\documentclass[aps,preprint,superscriptaddress,showpacs]{revtex4}
\usepackage[english]{babel}
\usepackage{braket}
\usepackage{slashbox}
\usepackage{epsfig}
\usepackage{amsfonts}
\usepackage{graphicx}
\usepackage{hyperref}
\usepackage{natbib}
\usepackage{amsthm}
\usepackage{amsmath}
\usepackage{times}
\usepackage{psfrag}
\usepackage{subfigure}
\usepackage{epstopdf}
\begin{document}

\title{Entrainment of a van der Pol-type circadian pacemaker to \textbf{daylight cycle}}

\author{\textbf{F. L. Tsafack Tayong}}
\affiliation{Fundamental Physics Laboratory, Physics of Complex System group,
Department of Physics, Faculty of
 Science, University of Douala, Box 24 157 Douala, Cameroon.}
\author{\textbf{R. Yamapi}}
\email[ryamapi@yahoo.fr]{(Corresponding author)}
\affiliation{Fundamental Physics Laboratory, Physics of Complex System group,
Department of Physics, Faculty of
 Science, University of Douala, Box 24 157 Douala, Cameroon.}
\author{\textbf{G. Filatrella}}
 \affiliation{Department  of Sciences and Technologies
\small and INFN Gruppo collegato Salerno, University of Sannio, Via Francesco De Sanctis,
I-82100 Benevento, Italy.}

\begin{abstract}
A van der Pol self sustained oscillator with higher order nonlinearity exhibits a rich dynamics, with multiple periodic attractors, and still the model allows analytical approximations.
Some of these properties can be conveniently exploited in the framework of circadian oscillations.
When interpreted as a biological oscillator that determines the alternation sleep/awake, the dynamic variable exhibits some interesting features that can be related to biological behavior.
We analyze in the paper the phenomenon of entrainment of van
 der Pol-type circadian pacemaker to daylight cycle. We determine the
  amplitude and frequency of the circadian model without natural forcing
  light, and find that the agreement of between analytical and numerical
  results hardly depend on the stiffness coefficient, $ \mu $ of circadian oscillations.
  It is shown that a practical and precise drive which imitates the effects of the
   conditions of natural light can be introduced in the system and analytically
    treated. Considering the effects of forcing light on the sleep aware cycle
    model, we find that a strong drive destroys the circadian oscillations and
    thus we confirm the importance of darkness for regular circadian oscillations.
    Moreover, we observe the reverse situation when we take into account the phase
    in the forcing light. For instance, for $\phi = \pi/2$ and the duration of daylight
    $D_L = 12h$, the phenomenon of quenching of circadian oscillations disappears.
    The comparison between analytical treatment and numerical simulations of effects of light is discussed.

\end{abstract}
\maketitle


\section{Introduction}
\label{Introduction}

The van der Pol oscillator is a paramount system to describe self\textbf{-sustained}
oscillations of artificial as well as biological systems.
No wonder, it has been employed to model one of the fundamental oscillations of human, and even of primitive bacterial, life: the circadian alternate phases of sleep and awake\cite{Kaur18}.
To such purpose the van der Pol  original version has been enriched with higher order nonlinearity, as well as with the coupling of other degrees of freedom to describe the circadian oscillations of various body functions, along with the external drive of the light \cite{Jewett98}.
At the same time, the mathematical model of self-sustained oscillations has been investigated to include several effects of higher order polynomial dissipation and to describe birhythmicity \cite{Yamapi10,Yamapi12}, or the response to external drives to describe forcing, noise\cite{Yamapi18}, as well as delayed perturbation \cite{Ghosh11,Biswas16,Guo18}.

The aim of this paper is to interpret the mathematical
developments in the context of the sleep / wake cycle which is a
biological phenomenon under active investigation with quantitative
methods which can lead to predict the circadian phase \cite{Stone20}.
The general objective is to seek analytical and numerical approximations which
describe the modified van der Pol type oscillator in the presence of a
 light stimulus and which may be useful for biological research.
  To do so we assume that the van der Pol osccilator simulatesTo do so we assume that the van der Pol osccilator simulates the stimulator whose biological name
  is \emph{supra-chiasmatic nucleus}  circadian stimulator
  responsible for the synchronization of the internal clock of the human
  organism located at the level of the brain \cite{Lit33} and
  considered  that the circadian oscillator is subjected to an interaction with the environment \cite{Smolen09} and trained to periodic light stimuli \cite{Kronauer99b}.
  In this work we propose to bring in a particular circadian light stimulus, which is extinguished for a certain time (the `` night '') and in the remaining part is sinusoidal in shape, to mimic the behavior of light  from the sun.
 We first investigate whether such behavior is able to lock the
 system and thus impose  circadian rhythm,
 and then we hilight  other interesting effects on the wake-sleep cycle.
  Thus, a purpose of the work is to modify the van der Pol oscillators
  to propose a specific realization of the nonlinear self-sustained
  oscillator that might fit the dynamic of the sleep/awake cycles with a
  peculiar form of the light stimulus. A second novelty consists in
  exploiting the mathematical treatability of the system to derive some analytical results.

The work is organized as follows: Section \ref{Cycle} presents the mathematical
formulation of the basic oscillator, and the
main analytical features of the undriven system  and for which we propose an analytical treatment in Sect. IIB. Section \ref{natural} proposes
a model for the light influence that are shown to be analytically manageable in Sect. IIIB, and
 whose effects are described in Sect. \ref{effects}.
Section \ref{conclusions} collects the conclusions.

\section{The  sleep awake cycle model and dynamical analysis}
\label{Cycle}
This Section describes the main features of the mathematical model to be investigated, in the interpretation of the sleep/awake circadian oscillations.

\subsection{The undriven oscillator to model the sleep awake cycle}
\label{Model}

The model used in our analysis for the sleep aware cycle model is the core self-sustained oscillator of the model for circadian oscillations determines \cite{Kronauer82,Kronauer90,Jewett99}, described by the following differential equation:
\begin{equation}
 \label{eq1}
   \left(\frac{12}{\pi}\right)^{2}\ddot{x}+\mu\left(-\frac{1}{3}-4x^{2}+ \frac{256}{15}x^{6}\right)\left(\frac{12}{\pi}\right)\dot{x}+\left(\frac{24}{\tau_{x}}\right)^{2}x=0.
\end{equation}
In the original
model \cite{Kronauer90}, $x$ is the
component of the human circadian pacemaker that
closely reflects the endogenous core body temperature
cycle, as measured during constant routine conditions\cite{czeisler-jewett1990}, $\tau_x$
represents the intrinsic period of the circadian oscillator
and $\mu$
 the stiffness of the oscillator that also regulates the nonlinear damping.
At variance with the model of Eq. (\ref{eq1}), the birhythmic model \cite{Yamapi10} is more general, as it contains also a term $\propto x^4$.
It is however important to underline that more refined models of the sleep/awake cycle are higher dimensional, inasmuch they contain also the processes that generate the circadian limit \cite{Indic05}.
In this work we employ the nonlinear oscillator (\ref{eq1}) as the most important oscillator that regulates the sleep/awake cycle, and we interpret some of the features of the oscillations of Eq.(\ref{eq1}) as indicative of the sleep/awake cycle.

\subsection{Analysis of the dynamics}

\begin{figure}
\begin{center}
\includegraphics[height=6.4cm,width=10.0cm]{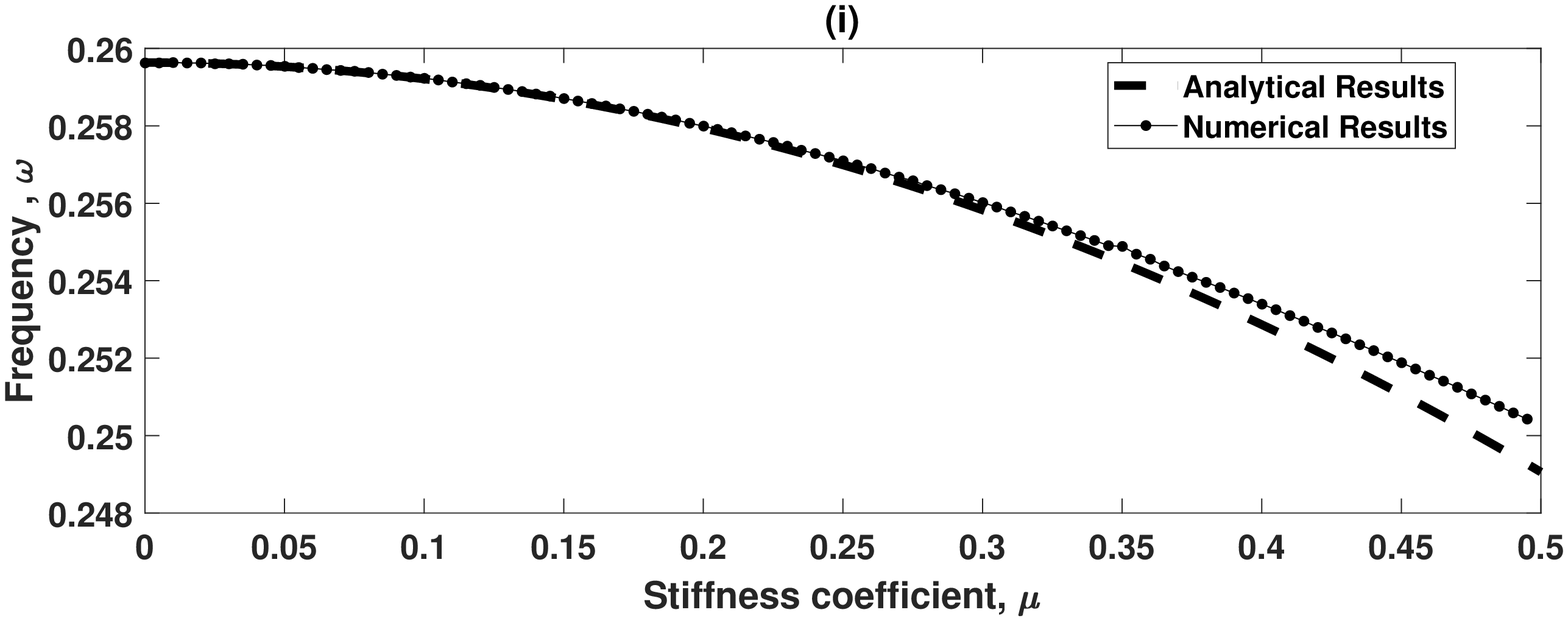}
\includegraphics[height=6.4cm,width=10.0cm]{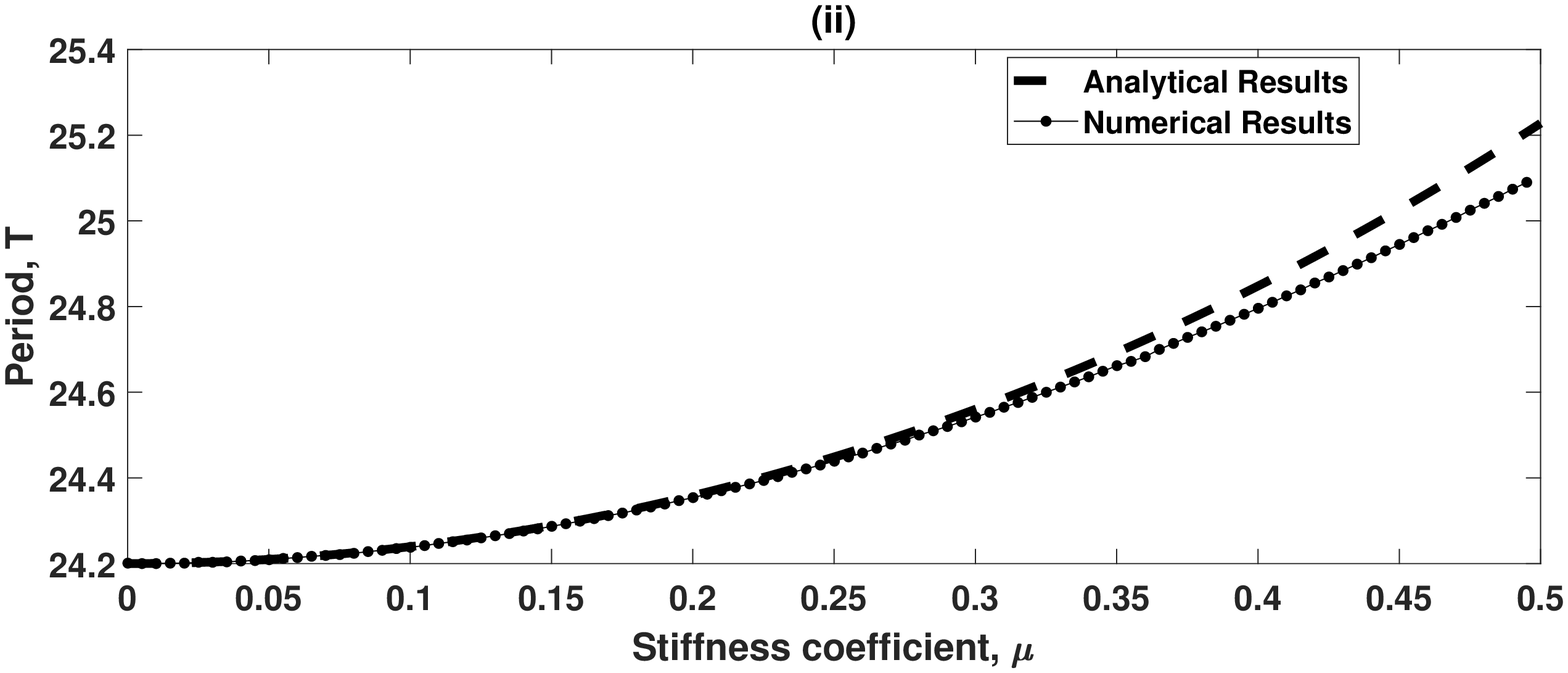}
\caption{\it Variation of the period $T$ (i) and of the frequency $\omega$ (ii) of
circadian oscillations versus the stiffness coefficient $\mu$.
These results are obtained using $Eq.(7)$.}
\label{fig1}
\end{center}
\end{figure}

We are going to determine analytically the amplitude and the frequency of the circadian oscillator model\cite{Kronauer82,Kronauer90,Jewett99}, for which we rewrite equation (\ref{eq1}) in the form below:
\begin{equation}
\label{eq2}
  \ddot{x}+\left(\frac{2\pi}{\tau_{x}} \right)^{2}x=-\mu\left(-\frac{1}{3}-4x^{2}+\frac{256}{15}x^{6} \right)\left(\frac{\pi}{12}\right)\dot{x}.
\end{equation}
Equation (\ref{eq2}) is the main equation of this
paper, whose analysis, also in the presence of a peculiar drive, is the subject of our investigation,
Lindsted perturbation method \cite{l1,l2} can be used to find the amplitudes and frequencies. Let us set $\tau =\omega t$, where $\omega $ is the unknown frequency, and to assume that the periodic solution of the differential Equation (\ref{eq2}) has the following approximate form:
 \begin{equation}
  \label{eq3}
  x(\tau)=x_{(0)}(\tau)+ \mu x_{(1)}(\tau)+ \mu^{2}x_{(2)}(\tau)+ ...
  \end{equation}
 where $ x_{(n)}(\tau)(n=0,1,2, · · · ) $ are functions of $ \tau $ of period $2\pi $.
 It is possible to write the frequency $ \omega $ in the form:
 \begin{equation}
 \label{eq4}
 \omega=\omega_{0}+ \mu \omega_{1}+ \mu^{2}\omega_{2}+ ...
  \end{equation}
  where $\omega_n(n = 0,1,...)$ are unknown constants.
Using the perturbation method, the periodic solution of Eq.(\ref{eq2}) can be approximated
by the following expression:
\begin{eqnarray}
\label{eq5}
x(t) =  && A\cos\omega t + \mu\frac{\pi}{\omega_{0}} \left[\left(\frac{73}{720}A^{7}-\frac{3}{96}A^{3}\right)\sin\omega t - \left(\frac{1}{40}A^{7}-\frac{1}{96}A^{3}\right)\sin3\omega t \right. \nonumber\\
  && \left.-\frac{1}{216}A^{7}\sin5\omega t-\frac{1}{2160}A^{7}\sin7\omega t \right] + O(\mu^{2}),
\end{eqnarray}
where the amplitude $A$  obeys to the following equation:
\begin{eqnarray}
\label{eq6}
A\left(\frac{1}{6} + \frac{1}{2}A^{2}-\frac{2}{3}A^{6} \right) = 0,
\end{eqnarray}
and the frequency $\omega$ given by:
\begin{equation}\label{eq7}
   \omega  =  \frac{2\pi}{\tau_{x}} - \frac{\pi^{2}}{\omega_{0}}\left[\frac{31}{10800}A^{12}-\frac{67}{17280}A^{8}-
   \frac{73}{51840}A^{6}+\frac{1}{1152}A^{4}+\frac{1}{2304}A^{2}\right]\mu^{2} +O(\mu^{3}).
\end{equation}
It is possible to  solve  Eq.(\ref{eq6}) to identify the conditions that give
  one root, and therefore one limit cycle, with amplitude $A \simeq 1.0$
  and the frequency $w\simeq 0.25745$ for the stiffness coefficient $\mu=0.23$.
 Figure \ref{fig1} shows the effects of
  stiffness coefficient $\mu$ on the behavior the frequency (see Fig.\ref{fig1}(a))
 and the period $T$ (see Fig.\ref{fig1}(b))
   of the circadian cycle.
The comparison between the analytical and the numerical results shows a nice agreement  for $\mu\le 0.35$, while for $\mu>0.35$ the analytical and numerical results diverge.
    These results are in agreement with the one obtained in 1995 by Choe and Czeiser \cite{ts}, who have estimated that the stiffness of human circadian pacemakers was between $0.005$ and $0.34$.
     Therefore, to use $\mu = 0.23$ for the van der Pol oscillator is  reasonable.
This oscillator is the basic element of the forced system, that is the subject of next Section.

\section{A model for the natural light forcing of the sleep-awake cycle}
\label{natural}

To properly insert a drive into van der Pol system with quintic dissipation to mimic the light/darkness effect, one should consider a function that is {\it blank} for some part of the period $24h$ (the night time), and displays a variable luminosity for the remaining part (the increase from the sunrise to noon, and then the decrease until dusk). A simple mathematical function that we propose is as follows:
\begin{eqnarray}
\label{forcing}
I(t) &=& \theta\left(D_L - t_{24}\right) I_0 \sin{\left(\omega  t_{24} +\phi \right)}, \\
\label{newperiod}
\omega &=& \frac{\pi}{D_L}, \nonumber\\
t_{24} &=& {\textrm mod} (t,24), \nonumber
\end{eqnarray}
here $I(t)$ is the forcing, $24h$ is the cycle period,
$D_L$ is the duration of the daylight, $\theta$ the Heaviside function.
The model so far introduced employs arbitrary units.
However, we find it convenient to have the time normalized to $1h$, so that a day corresponds to $t=24$, to facilitate an intuitive interpretation of the results.
Thus, $D_L=12$ denotes the equinox duration of the day (neglecting twilight effects).
The corresponding angular velocity reads $\omega=\pi/12$, the phase $\phi$ mimics a shift of the daylight respect to the cycle.
This shift might be relevant for the transient analysis, or to describe a sudden shift respect to the daylight variation, as can occur in intercontinental flights, where sudden time shift $\Delta t$ can appear that induces a change in the phase $\Delta \phi = \omega \Delta t = \pi \Delta t/D_L$.
Implicitly, we are assuming that on the scale of the oscillations of the variable “x” (some proxy of circadian biological oscillations such as the body temperature), the frequency of the irradiation is too fast to have an effect, and only its intensity is relevant.

\begin{figure}[h!]
    \begin{center}
\includegraphics[height=6.4cm,width=10.0cm]{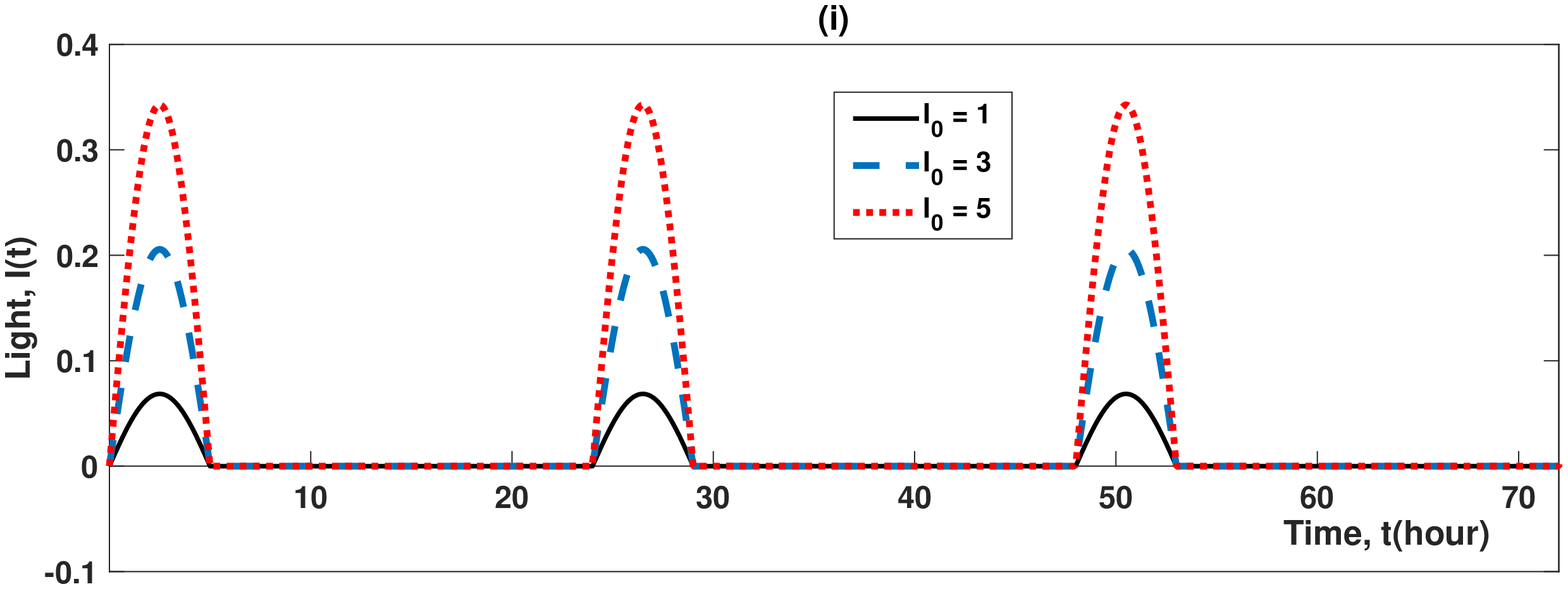}
\includegraphics[height=6.4cm,width=10.0cm]{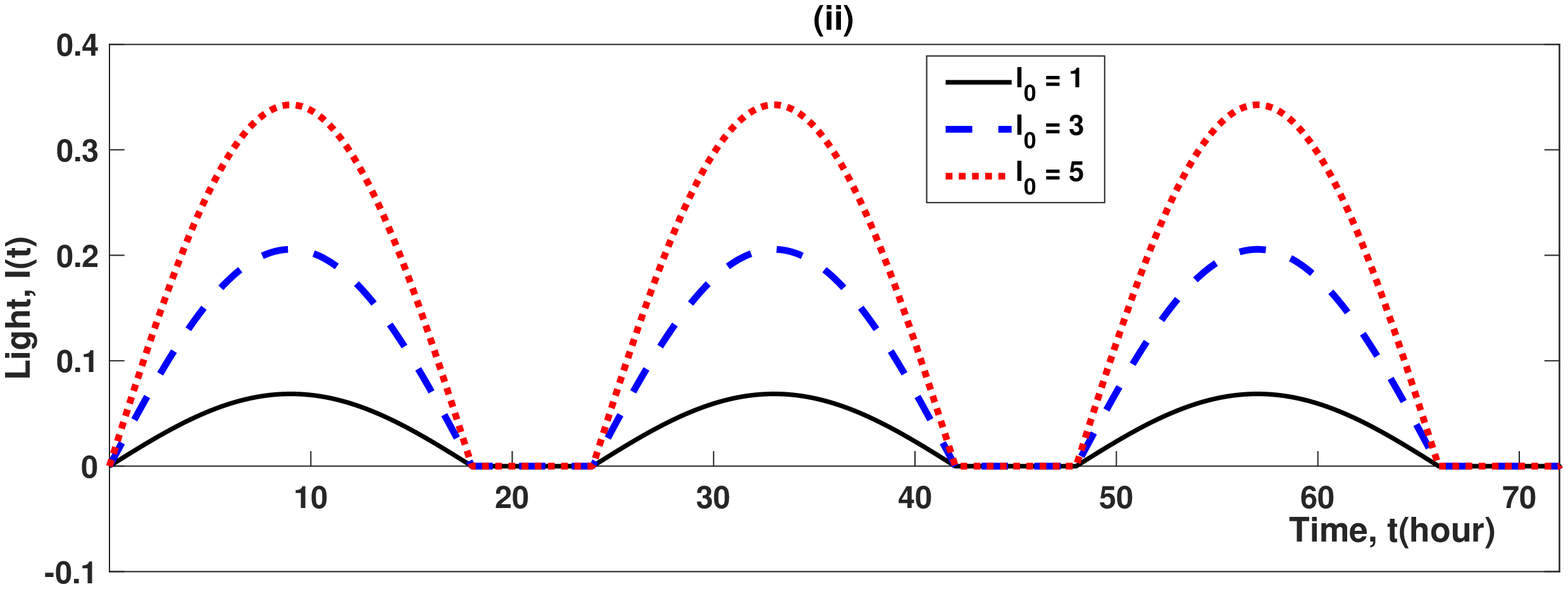}
    \end{center}
   \caption{\it Oscillations of the natural light $I(t)$ in three days with several values of the amplitude $I_0$ and of the duration of the daytime $D_L$:
    (i):$ D_{L}= 5h $ and  (ii): $D_{L}= 18h$.  These results are obtained using $Eq.(8)$.     }
   \label{fig2}
\end{figure}

Figure \ref{fig2} shows the variation of the light (\ref{forcing}) in three days with two different values of the amplitude $I_0$ of the light and for $D_L =5h$ (i) and $D_L= 18h$ (ii), consistent results are obtained for other values of the daylight duration.
It is important to underline that the Heaviside function multiplies a sinusoidal function, thus altering the drive that is not anymore sinusoidal as in \cite{Yamapi18}.
However, the analytic estimation for the purely sinusoidal drive can still be a guideline for the system, as will be shown below.

\section{The effects of the light on the sleep aware cycle model}
\label{effects}

In this Section, we analyze the influence of natural light parameters on the dynamic behavior of the Jewett et al. model (\ref{eq2})\cite{Jewett99}.
At the core of the Jewett et al. model is the following modified driven van der Pol (mvdP) differential equation:
\begin{equation}
 \label{eq10}
   \left(\frac{12}{\pi}\right)^{2}\ddot{x}+\mu\left(-\frac{1}{3}-4x^{2}+ \frac{256}{15}x^{6}\right)\left(\frac{12}{\pi}\right)\dot{x}+\left(\frac{24}{\tau_{x}}\right)^{2}x =I(t).
\end{equation}
 It will therefore be a question to check weather  the duration of the daylight $D_L$ and of the intensity $I_0$  of the forcing influences the sleep-wake cycle.
 In other words, it is relevant to study  the effects of $I_0$ and $D_L$ on the period and the amplitude of the drive of the nonlinear oscillator.
We will proceed to an analytical investigation and we complete
the analysis with numerical simulations.

\subsection{Analytical investigations}

\begin{figure}
\begin{center}
\includegraphics[height=6.40cm,width=10.0cm]{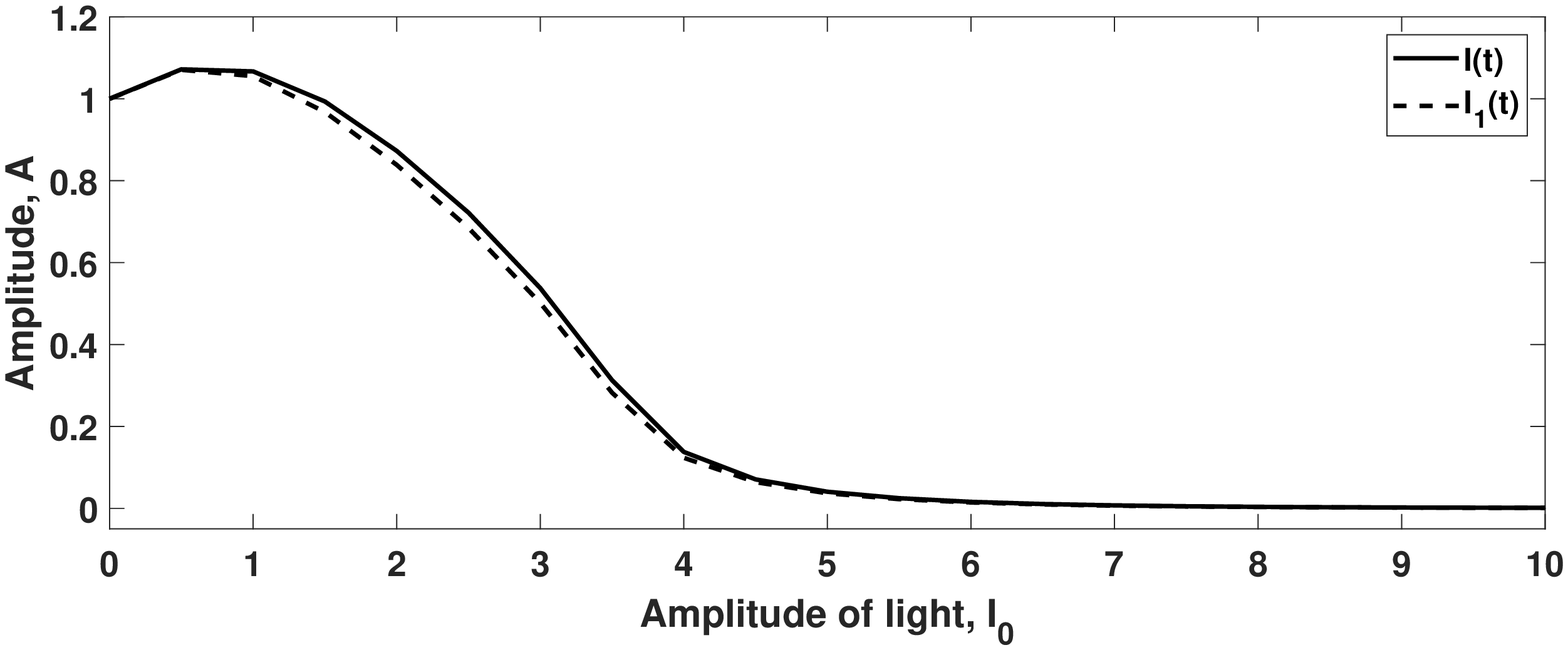}
\caption{\it   Comparison of the variation of the amplitude of sleep as a function of $ I_ {0}$ 
for the forcing $I(t)$ (see Eq.(8)) and its Fourier series at 
first order $ I_ {1} (t)$(see Eq.(13)). Numerical method for $ I_ {0}(t)$ 
 and $ S_ {1} (t) $ ( $D_{L}= 12h $). The other parameters are: $ \mu = 0.23$, $\tau_{x} = 24.2 $.}
\label{fig12}
\end{center}
\end{figure}

It is possible to gain an analytical insight of the system (\ref{eq10})
under the influence of the peculiar shape of the light (\ref{forcing}) using the Fourier series of the signal.
As the signal is periodic, the Fourier expansion (to the order $N$) is promptly written in the form:
\begin{eqnarray}
\label{transformation3}
I_n (t) =\frac{a_{0}}{2} +  \sum\limits_{\substack{ i=1}}^{n}{( a_{i}\cos(i\omega t)+ b_{i}\sin(i\omega t))},
\end{eqnarray}
with the coefficients given by:
\begin{subequations} \label{E:11}
  \begin{align}
     a_{0} & = \frac{\omega}{\pi} \int_0 ^{\frac{2\pi}{\omega}} I(t)dt,   \label{E:11a}\\
a_{n} &= \frac{\omega}{\pi} \int_0 ^{\frac{2\pi}{\omega}} I(t)\cos(n\omega t) dt,   \label{E:11b}\\
 b_{n} &= \frac{\omega}{\pi} \int_0 ^{\frac{2\pi}{\omega}} I(t)\sin(n\omega t) dt. \label{E:11c}
  \end{align}
\end{subequations}
Inserting the light (\ref{forcing}) it is possible to retrieve the coefficients:
\begin{equation}
\label{eq12}
    a_{0} = \frac{2I_{0}}{\pi}\cos\phi  , \quad     a_{1} = \frac{I_{0}}{2} \cos\phi \quad ,\quad b_{1} = \frac{I_{0}}{2} \sin\phi  .
\end{equation}
For $  n=1 $, the first Fourier component reads:
\begin{equation}
\label{E:13}
I_1(t) =\frac{ I_{0}}{\pi}\cos\phi   + \frac{I_{0}}{2}\sin(\omega t + \phi)
\end{equation}
Figure \ref{fig12} compares the amplitude of the system oscillations
 as a function of the original light fluctuations $I(t)$ of Eq.(\ref{newperiod})
 and of the first Fourier component $I_1(t)$ estimated through Eq. (\ref{E:13}).
The results show that the full model and the first component
 produce almost identical oscillations, and therefore it seems
 that just the first Fourier component captures most of the effect.
Therefore, we can now proceed for analytical investigations of the model (9) by consider
the mvdP model driven by the first component:
\begin{equation}
 \label{eq14}
\ddot{x}+\omega_{0}^{2}x = -\mu\left(-\frac{1}{3}-4x^{2}+ \frac{256}{15}x^{6}\right)\left(\frac{\pi}{12}\right)\dot{x} + I_{1a}   + I_{1b}\sin(\omega t + \phi),
\end{equation}
 with $ I_{1a} = \left(\frac{\pi}{12}\right)^{2}\frac{I_{0}}{\pi}\cos(\phi) $ and $ I_{1b} = \left(\frac{\pi}{12}\right)^{2} \frac{I_{0}}{2} $.\\
For this purpose, let us set the solution of equation (14) in the form:
\begin{equation}
 \label{eq15}
 x(t) =A_{1}cos\Omega t + A_{2}sin\Omega t+A_3=A\cos(\omega t-\phi_1)+A_3\textbf{•},
\end{equation}
where $ A=\sqrt{A_{1}^{2}+A_{2}^{2}} $   and  $\phi_1= \arctan(\frac{A_{2}}{A_{1}})$.\\
Inserting Eq. (\ref{eq15}) in Eq. (\ref{eq14}) and equating the cosine and sine terms separately, gives:

\footnotesize
\begin{subequations}
\label{E:18}
  \begin{align}
& I_{1b}\cos\phi = \left( \Omega^{2}-\omega_{0}^{2} \right)A_{1}+\left( \frac{5}{64}c\Omega A^{6}+\frac{15}{8}c\Omega A_{3}^{2}A^{4} + ( \frac{1}{4}b\Omega + \frac{15}{4}c\Omega A_{3}^{4})A^{2} + (c A_{3}^{4} + b)\Omega A_{3}^{2} + a\Omega \right)A_{2} ,\label{E:18a}\\
&I_{1b}\sin\phi =  -\left( \frac{5}{64}c\Omega A^{6}+\frac{15}{8}c\Omega A_{3}^{2}A^{4} + ( \frac{1}{4}b\Omega + \frac{15}{4}c\Omega A_{3}^{4})A^{2} + (c A_{3}^{4} + b)\Omega A_{3}^{2} + a\Omega \right)A_{1}  + \left( \Omega^{2}-\omega_{0}^{2} \right)A_{2} , \label{E:18b}\\
&A_{3}  = \frac{I_{1a}}{ \omega_{0}^{2}},\label{E:18c}
  \end{align}
\end{subequations} \normalsize
 with  $ a = \mu\frac{\pi}{36}$, $ b = \mu\frac{\pi}{3}$, $ c = -\mu\frac{256\pi}{180}$.\\
The amplitude $A$ satisfies the following equation:
\footnotesize \begin{equation}
 \left( \frac{5}{64}c\Omega A^{6}+\frac{15}{4}c\Omega A_{3}^{2}A^{4} + ( \frac{1}{4}b\Omega + \frac{15}{4}A_{3}^{4})A^{2} + c\Omega A_{3}^{6} + b\Omega A_{3}^{2} + a\Omega \right)^{2}A^{2} + ( \Omega^{2}-\omega_{0}^{2})^{2}A^{2} - I_{1b}^{2}=0.
  \label{eq19}
 \end{equation} \normalsize
Having derived the above equation, we propose in the following Section
 the numerical approach to handle the model (17). The results
 thus obtained will be compared to numerical simulations of Eq.(9).

 \subsection{Numerical simulations and discussions}
\subsubsection{Case of zero initial phase of the light}

\begin{figure}
\begin{center}
\includegraphics[height=6.40cm,width=10.0cm]{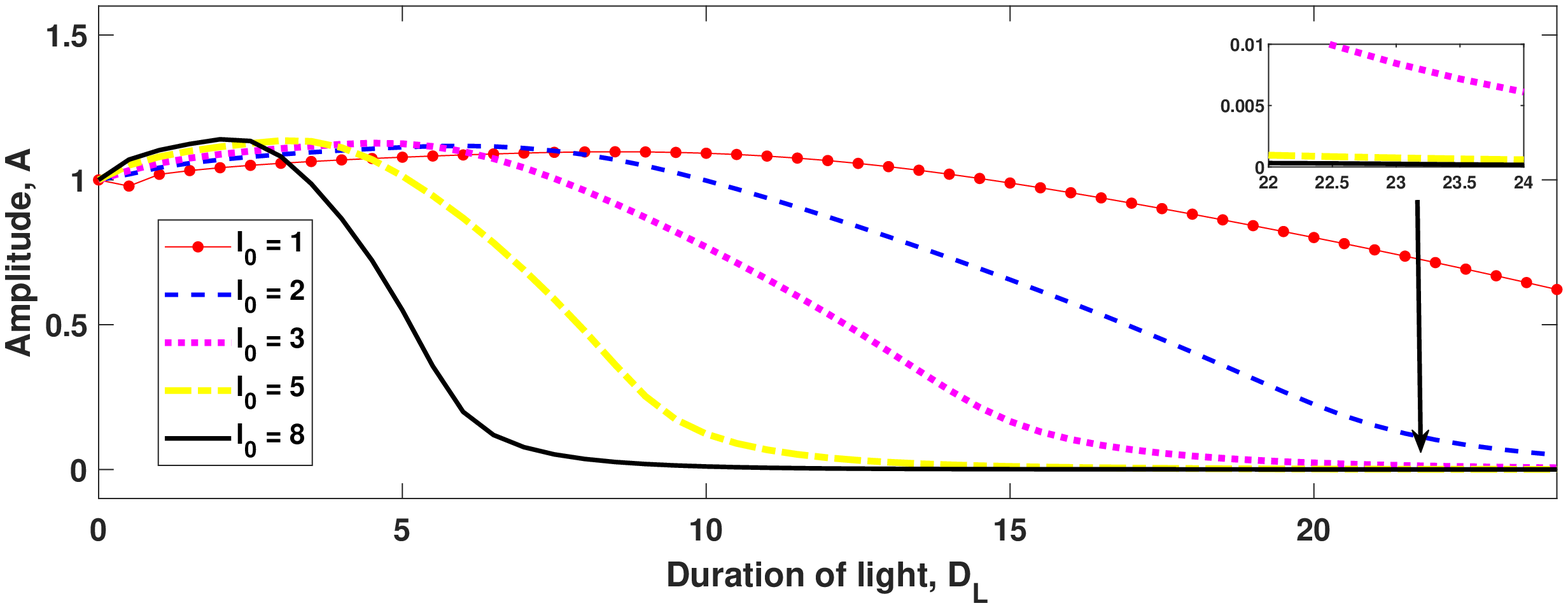}
\caption{\it Effects of the amplitude $I_0$ on the variation of the analytical amplitude $A$ of
circadian oscillations versus the
 duration of daylight $D_{L}$.
The other parameters are defined in Fig. \ref{fig12} and these results are obtained using $Eq.(17)$.}
\label{fig3}
\end{center}
\end{figure}

\begin{figure}
\begin{center}
\includegraphics[height=6.40cm,width=10.0cm]{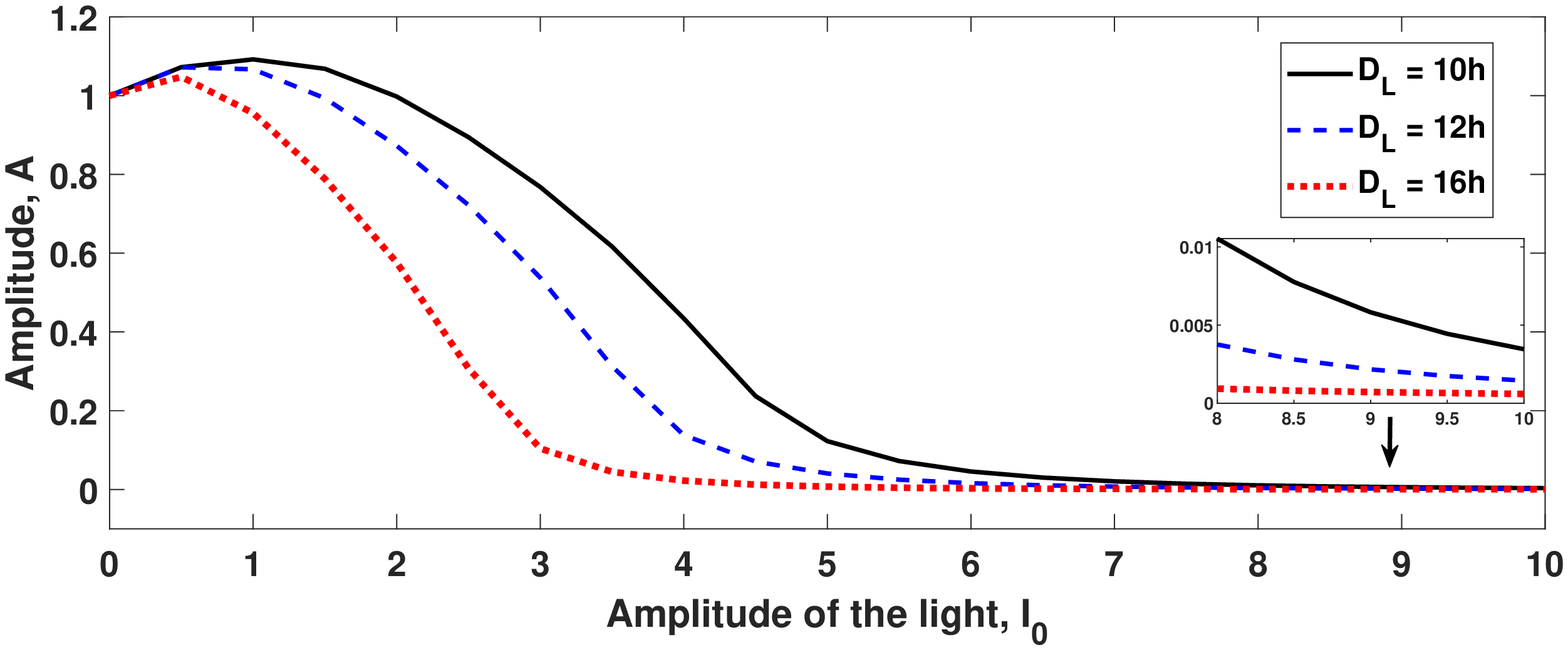}
\caption{\it Effects of the duration of daylight $D_L$ on the variation of the analytical
amplitude $A$ of oscillations versus the intensity of the light $I_0$.
 The other parameters are defined in Fig. \ref{fig12} and these results are obtained using $Eq.(17)$.}
\label{fig4}
\end{center}
\end{figure}

\begin{figure}[htbp]
\begin{center}
\includegraphics[height=6.40cm,width=10.0cm]{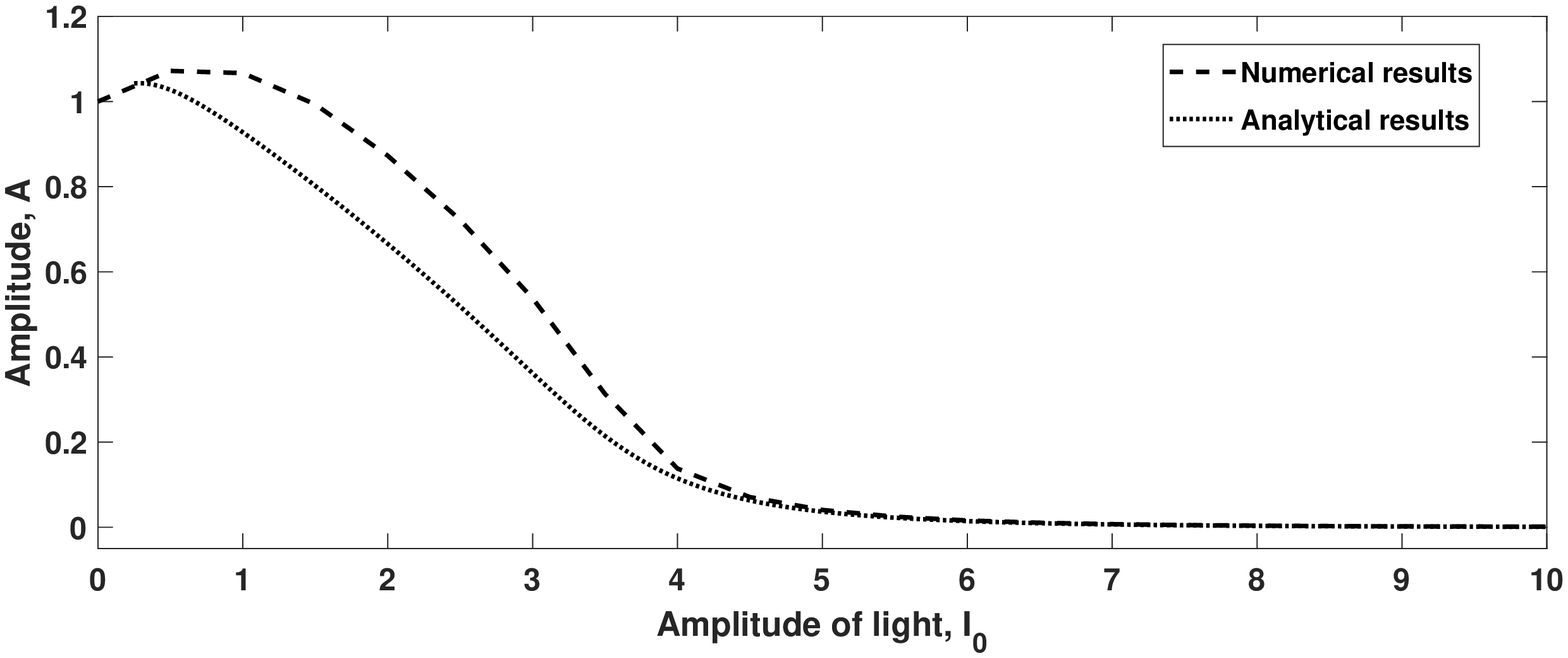}
\caption{\it Comparison of analytical and numerical results of the amplitude $A$ 
as a function of $ I_{0} $ for the Fourier series at first order  $ I_{1} (t)$ 
with $D_{L}= 12h $. The other parameters are defined in Fig. \ref{fig3}.
(Analytical and numerical results are obtained from Eq.(17) and Eqs.(9,13), respectively).}
\label{fig13}
\end{center}
\end{figure}

Let us assume that the initial phase of the natural light reads $\phi=0.0$.
The duration of daylight $D_L$ is chosen between $10$ and $20$ hours,
with an average duration, including dawn and dusk, of about $15$ hours.
Indeed, we can note for example that in the Northern hemisphere,
the summer days are longer while the winter days are shorter, and
the twilight lasts about one hour.
Figures \ref{fig3} and \ref{fig4} show the effects of the amplitude
$I_0$ and the duration of the daylight $D_L$, respectively, on the
temporal evolution $x(t)$ of the mvdP model under the action of light.
It appears that the duration of the daylight $D_L$ influences the behavior of the mvdP. To be concrete, the figure might represent the effects of the daylight from darkness for almost all day ($D_L \simeq 0$, e.g., summer at the south polar circle) to longer light hours (e.g., as one moves North during summer) the case for all-day light ($D_L \simeq 24$, e.g. summer at the north polar circle).
The resulting behavior is quite peculiar.
The oscillator displays a finite, relatively large amplitude only
if the light duration is {\it short} enough, and disappears as the light condition extends. This is indicative of possible sleep disturbances if the subject is exposed to darkness for a too short period.
It is also noticeable that a too strong drive destroys the circadian oscillations,
 again confirming the importance of darkness for regular circadian oscillations.
Thus the duration of the light and the light intensity play a role in the amplitude of sleep;  in fact, the amplitude of sleep decreases as $D_L$ and $I_0$ increase.  It is therefore clear that when one is exposed to strong light intensities for a long time, the desire to sleep decreases considerably until it tends to be canceled.  In other words, it is hardly possible to fall asleep under strong illumination for a long time.
Figure \ref{fig13} shows the partial adherence
between the numerical and analytical results of the effect of the light amplitude
$I_0$ on sleep for daylight $ D_{L}=12h $.

\begin{figure}
\begin{center}
\includegraphics[height=6.40cm,width=10.0cm]{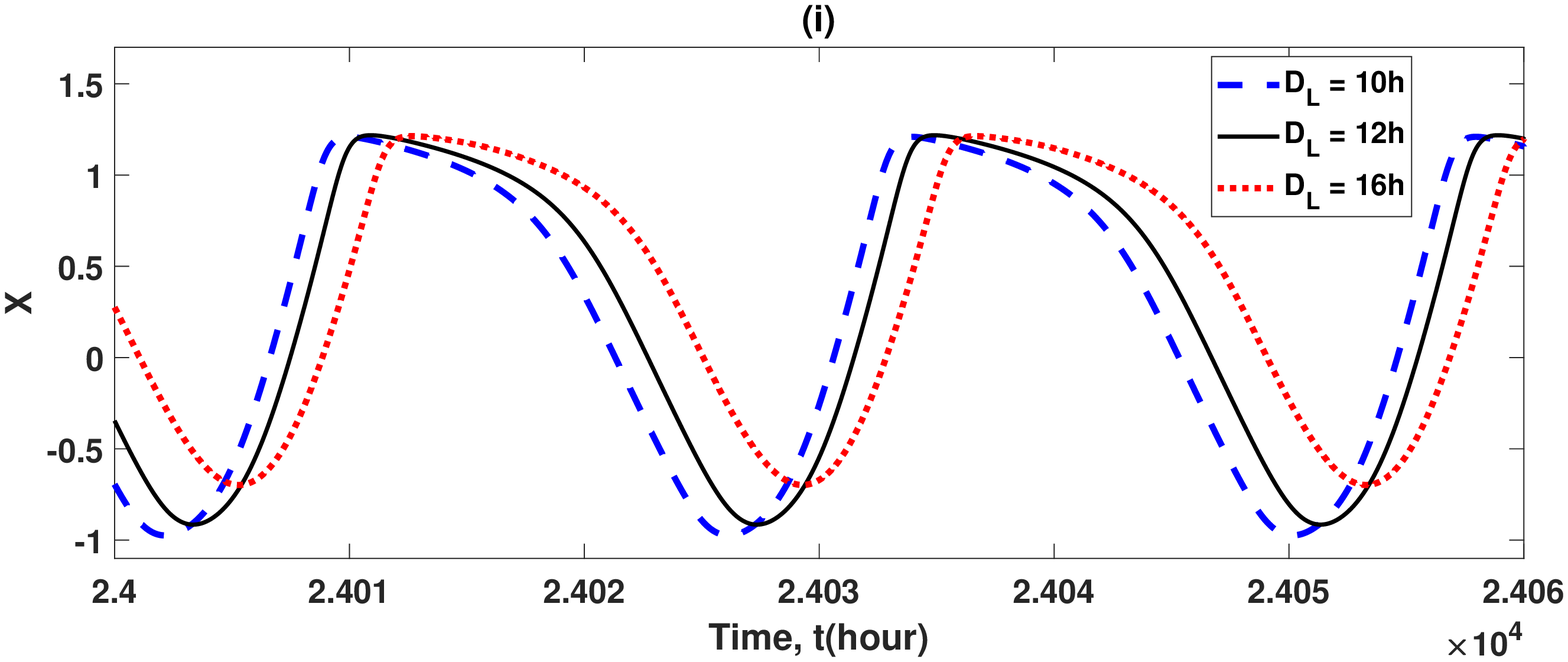}
\includegraphics[height=6.40cm,width=10.0cm]{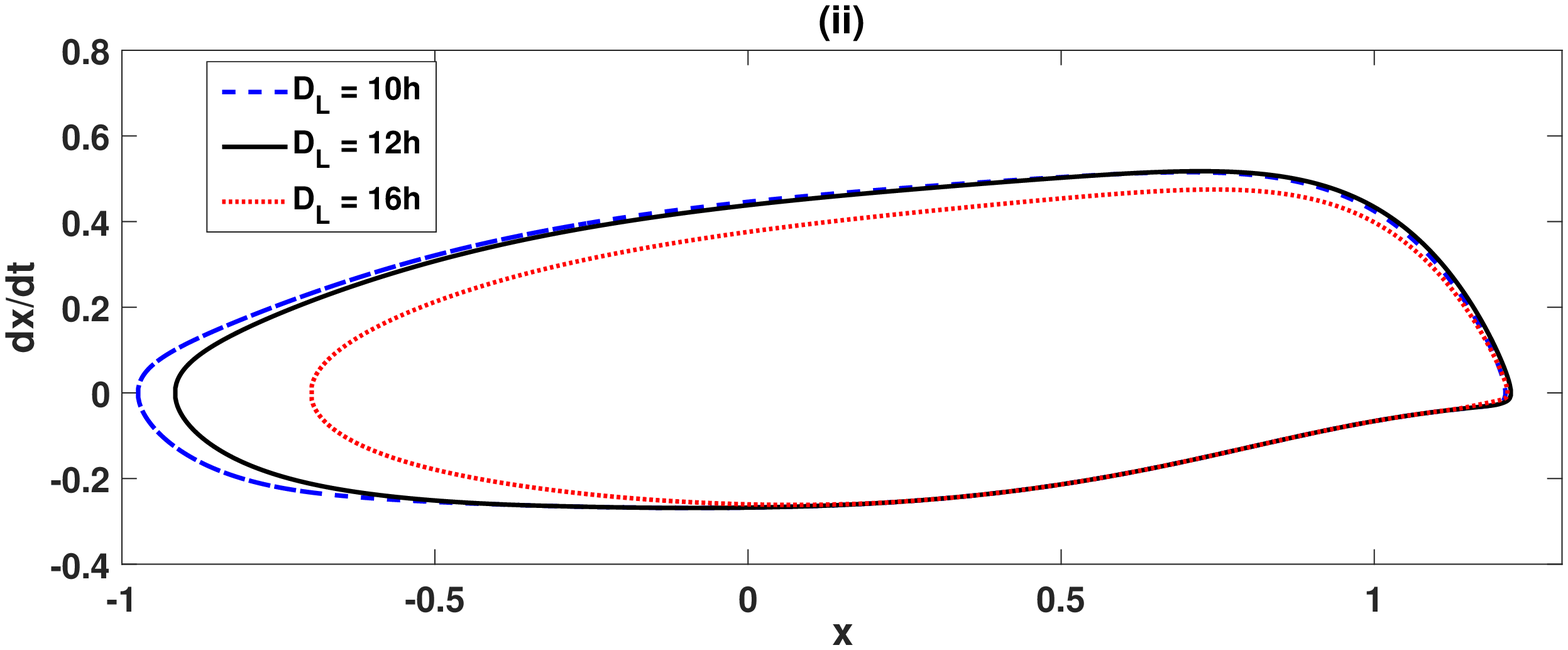}
\caption{\it Temporal evolution of $x(t)$ (i)  and
phase portrait (ii) for several different values of the duration of daylight:
 $ D_{L} = 10 h $ (solid line), $ D_{L} = 12 h $ (dashed line), and $ D_{L} = 16 h $ (dotted line) for  $ I_{0} = 1.00$.
The other parameters are defined in Fig. \ref{fig12} and results are obtained using Eqs.$(9,19)$.}
\label{fig5}
\end{center}
\end{figure}

\begin{figure}
\begin{center}
\includegraphics[height=6.40cm,width=10.0cm]{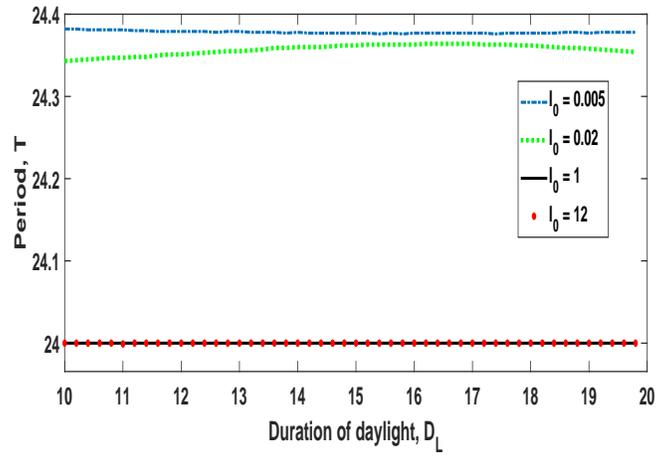}
\caption{\it Variation of the period $T$ of circadian oscillations versus the duration of daylight $D_{L}$
for some several different values of the intensity $I_{0}$ of the light.
The other parameters are defined in Fig. \ref{fig12} and results are obtained using Eqs.(9,13).}
\label{fig6}
\end{center}
\end{figure}

\begin{figure}
\begin{center}
   \includegraphics[height=6.40cm,width=10.0cm]{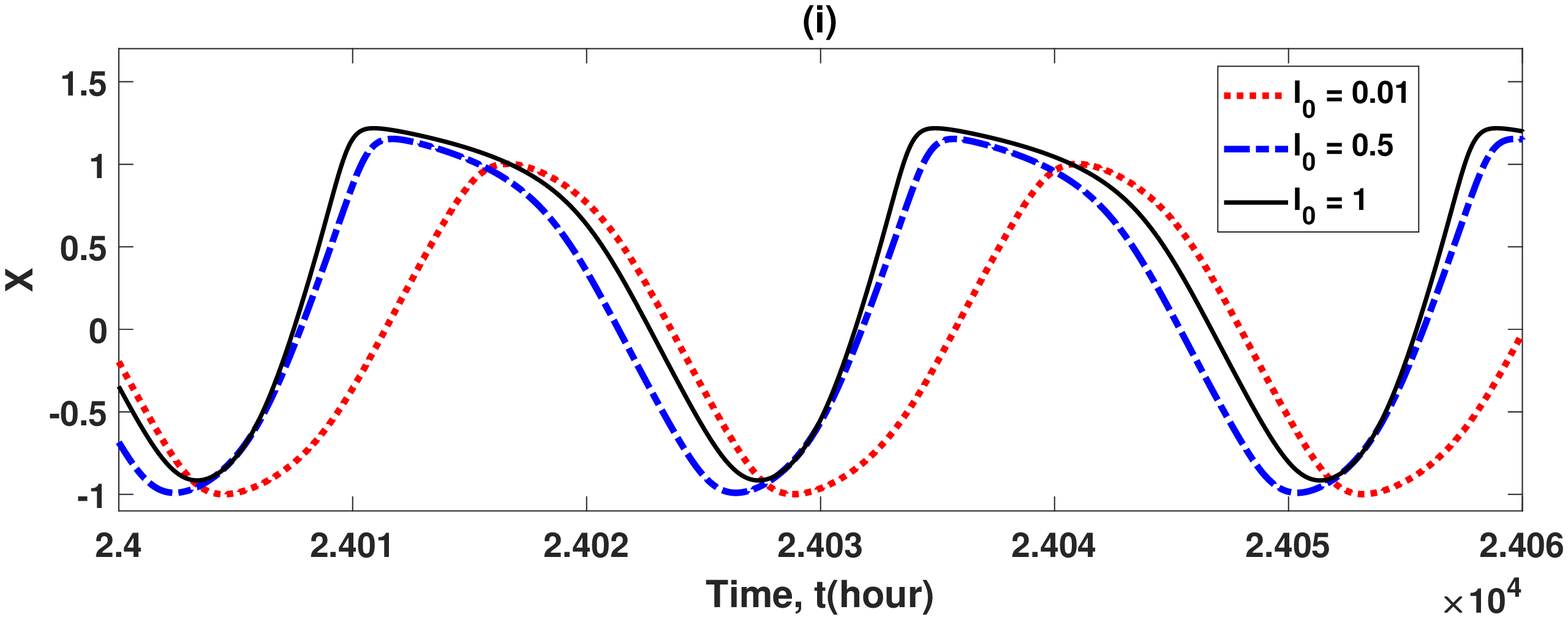}
   \includegraphics[height=6.40cm,width=10.0cm]{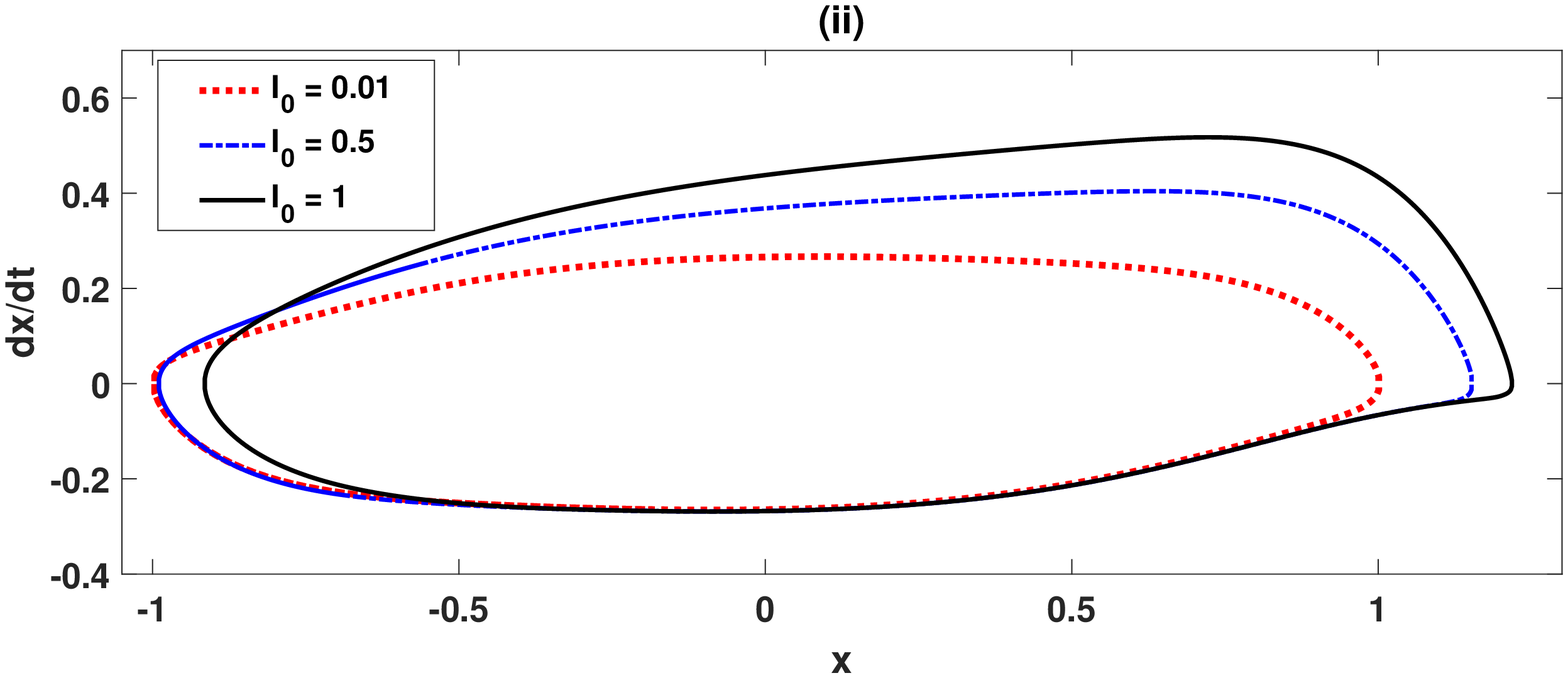}
\caption{\it Effects of the intensity  of the light $ I_{0} $ on the
temporal evolution of $x(t)$ (i) and phase portrait (ii)
with $ D_{L} = 12h $. The other parameters are defined in Fig. \ref{fig12}
and results are obtained using Eqs.(9,13).}
\label{fig7}
\end{center}
\end{figure}

Figure \ref{fig5} shows the effect of the amplitude $I_0$ on the temporal evolution $x(t)$ of the mvdP model under the action of light. It is evident that for the nonlinear oscillator (\ref{eq3}), the period depends upon the amplitude of the drive.
Figure \ref{fig6} shows the variation of the period $T$
versus the daylight $ D_{L} $ for several different values of the amplitude $I_0$. It appears that the period of oscillations changes automaticaly depending for the value of the amplitude $I_0$ (low $I_0$ or great $I_0$). This effects  are shown on Fig. \ref{fig7} where we plot the evolution of x(t)
and the phase shift for several different values of the amplitude $I_0$ with the daylight $ D_{L} = 12h $.

\begin{figure}[h!]
    \begin{center}
   \includegraphics[height=6.40cm,width=10.0cm]{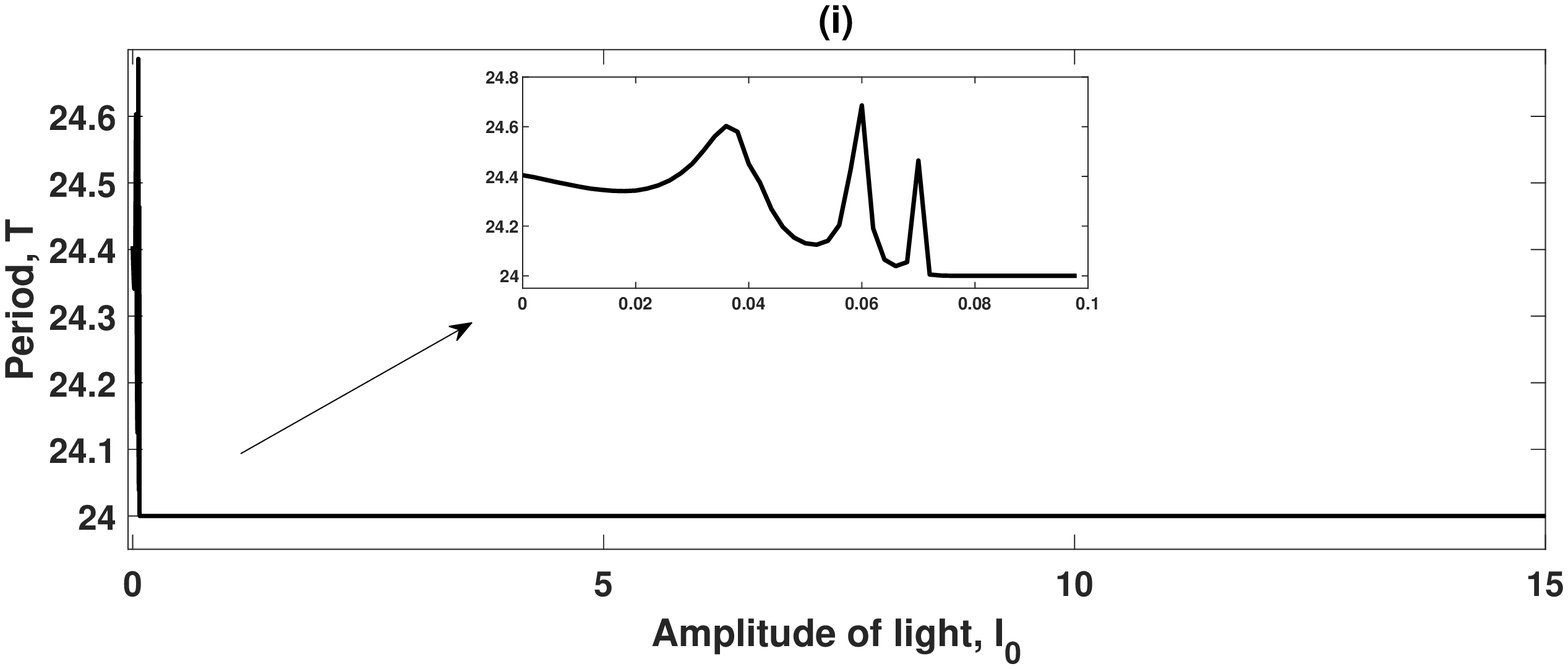}
   \includegraphics[height=6.40cm,width=10.0cm]{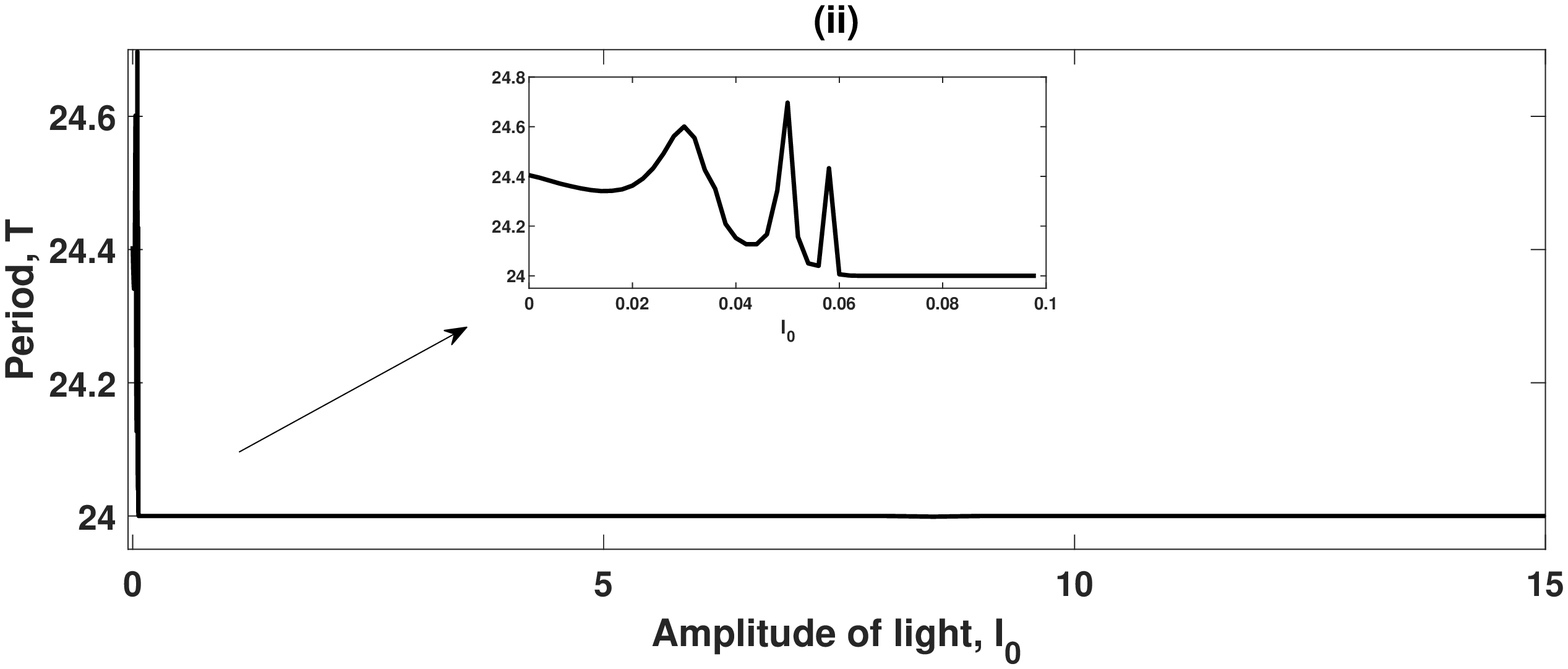}
    \end{center}
   \caption{\it Effects of the duration of daylight $D_{L}$ on the variation of the period $T$ of circadian oscillations versus the amplitude of the light $I_{0}$. $ D_{L}= 10h $ (i) and $D_{L}= 16h $ (ii).  The other parameters are defined in Fig. \ref{fig12} and results are obtained using Eqs.(9,13).}
   \label{fig8}
   \end{figure}

\begin{figure}
\begin{center}
\includegraphics[height=6.40cm,width=10.0cm]{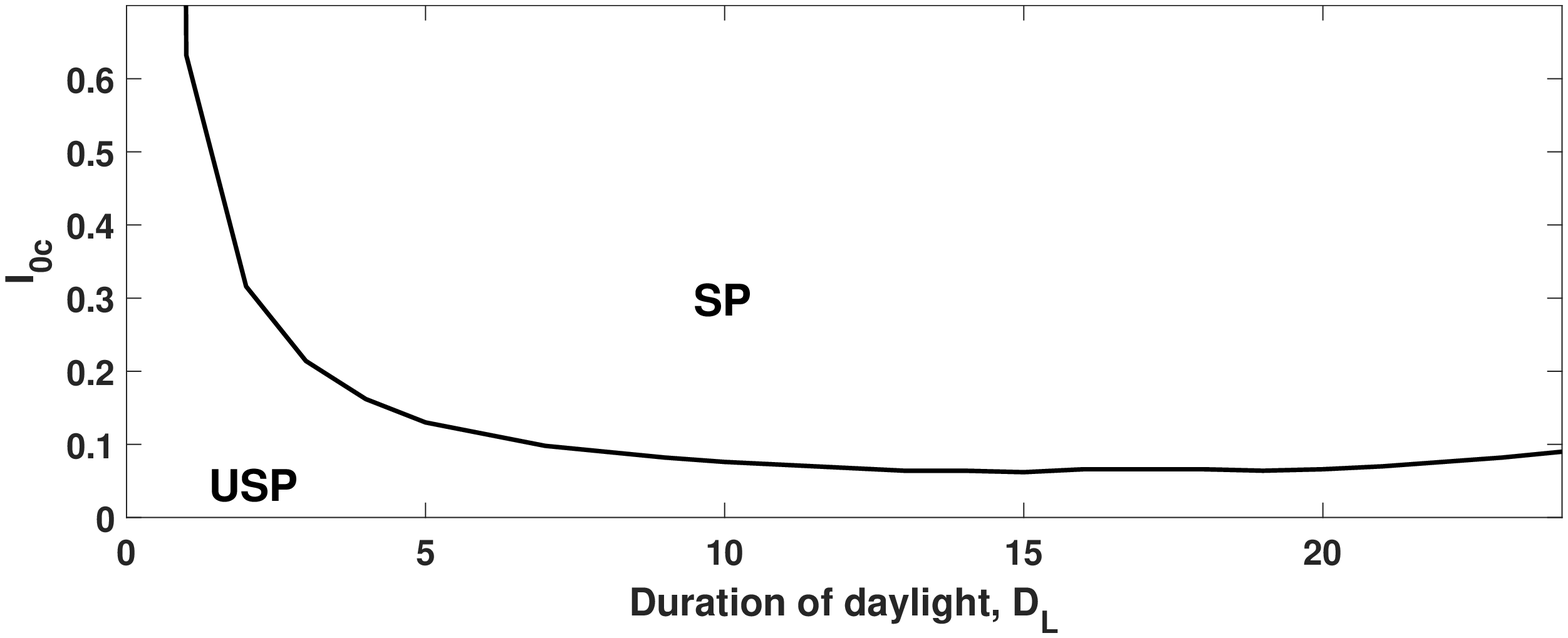}
\caption{\it Boundary in the $(I_0,D_L)$ plane between the region of synchronized period (SP) and the
region of unsynchronized period (USP).
The other parameters are defined in Fig. \ref{fig12} and results are obtained using Eqs.(9,13).
}
\label{fig9}
\end{center}
\end{figure}

In Figure \ref{fig8}, we present the variation of the period $T$ of circadian
oscillations versus the  amplitude $I_{0}$ of the light for several different values of the duration of daylight $ D_{L} $. It appears that when the amplitude $I_0$ increases from zero, the light does not induces in the mvdP model oscillations at $24h$ below a critical value $I_{0c}$ at which synchronization
(that is, the period of the mvdP oscillator matches the period of the drive within numerical approximation)
 of the mvdP model and the light  appears.
Let us note that this critical value $I_{0c}$ depends on the system parameters,
such as the duration of the daylight. Table \ref{tab1}  shows how $I_{0c}$
depends on the duration of the daylight $D_L$.
It appears that the increase of the duration of the light $ D_{L} $ leads to the decrease in the intensity $ I_{0c} $, moreover to have a large critical value, a value of the duration of  very small light ($ D_{L} \leq 0.001 $).
 In other words, depending on the season, it emerges that a critical intensity is needed in the summer is about $ I_{0c} = 0.064 $ and in the winter $ I_{0c} = 0.2 $ to have a standby cycle /  24h sleep.
 Fig.  \ref{fig9} shows in the $(I_{0c},D_L)$ plane, the boundary between
 the region where the light synchronizes the system to $24h$ (SP: $T=24h$) and
 the region where the light does not synchronize the system to 24h (USP:$T\neq 24h$).

\begin{figure}
\begin{center}
 \includegraphics[height=6.40cm,width=10.0cm]{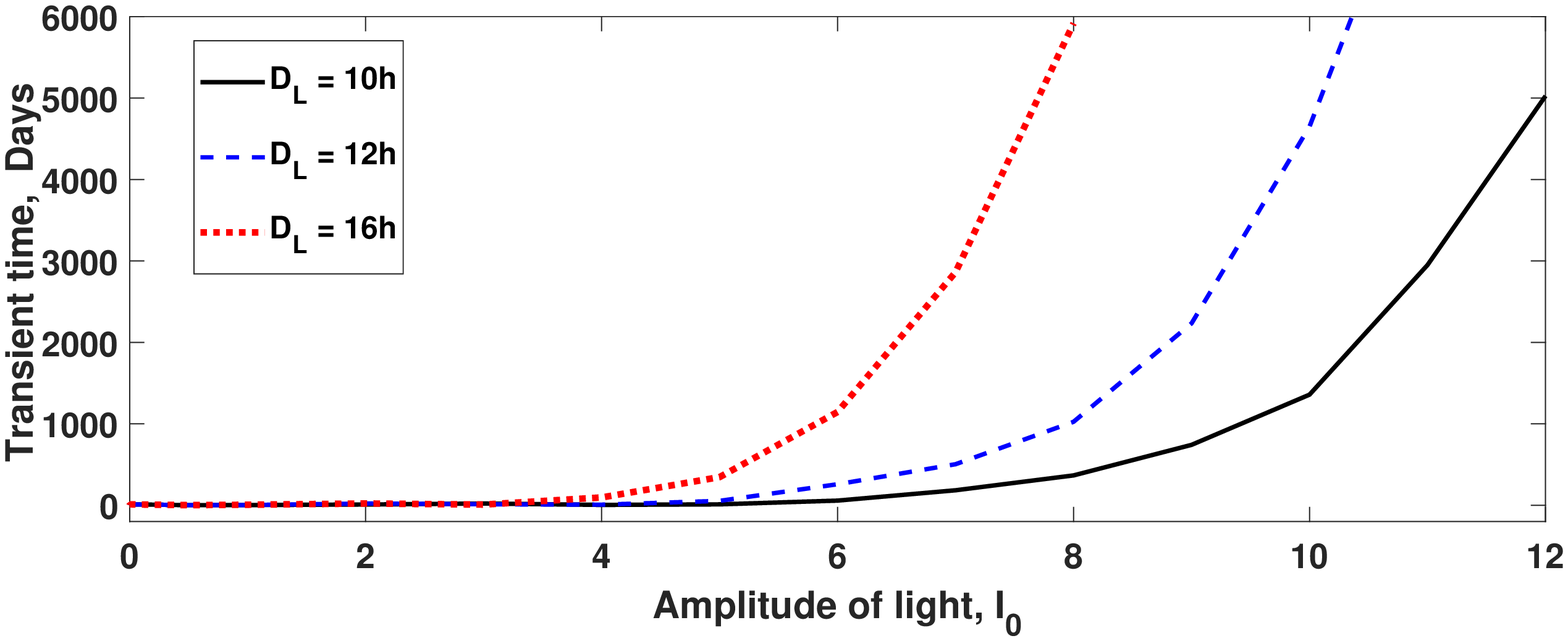}
\caption{\it Effects of the duration of dayligth $D_L$ on
the transient time versus $I_{0}$.
     The other parameters are defined in Fig. \ref{fig12}
     and results are obtained using Eqs.(9,13).}
\label{fig10}
\end{center}
\end{figure}

      \begin{table}[h!]
        \begin{center}
           \begin{tabular}[b]{|c|l|l|l|l|l|l|l|l|l|l|l|l|l|l|}
                        \hline
$D_L $ & 0.001 & 1 & 2 &  4  & 6  & 8 & 10 & 12  & 14 & 16 & 18 & 20 & 22 & 24 \\
                         \hline
  $  I_{oc} $ & 15 & 0.632 & 0.316 & 0.162 & 0.114 & 0.090 & 0.076 & 0.068 & 0.064 & 0.066 & 0.066 & 0.066 & 0.076 & 0.09 \\                                       \hline
                    \end{tabular}
        \caption{\it Boundary between synchronization states  and unsynchronization states in the $(I_0,D_L)$ plane ($SP$ = Synchronized period / $USP$ = Unynchronized period). 
        These results are obtained by numerical simulations of $Eqs.(9,13)$ and shown on figure $11$.}
          \label{tab1}
        \end{center}
      \end{table}
Given that the period of the undriven oscillator is not $24$ hours, a certain time, called transit time, is needed, during which the forcing brings the oscillator to a period of $24$ hours.
Table \ref{tab2} gives more information on the duration of the transit time as a function of $ I_0 $ and $ D_{L} $.
It appears that  the increase in the duration of the light $ D_{L} $ and of the intensity $ I_0 $ leads to an increase
 in the duration of the transition time, from one day to several months (Fig. \ref{fig10}) .

  \begin{table}[h!]
        \begin{center}
           \begin{tabular}[b]{|c|l|l|l|l|l|l|l|l|l|l|l|l|}
                        \hline
                       \backslashbox{$ DL$}{$I_{0} $}   & 0.0025 & 0.005  & 0.5 & 1 & 2 & 3 & 4 & 5 & 6 & 7 & 8   \\
                         \hline
10  & 11.46 & 11.04  & 1.33 & 2.42 & 9.33 & 21.37 & 4.33  & 11.33 & 57.29 & 184.17 & 366.67   \\
\hline
12  & 11.46 & 11.042  & 1.37  & 2.50    & 21.25  & 12.5   & 7.08  & 54.17 & 258.33 & 504.17& 600.6     \\
\hline
16  & 11.46 & 11.04 & 1.46 & 6.67 & 21.87 & 8.50    & 95.83  & 345.83 & 1145.83 & 2862.50  &  5916.67 \\
                        \hline
                    \end{tabular}
        \caption{\it Transient time(days) for several different values of $I_{0} $ and $ DL$. 
        These results are obtained by numerical simulations of $Eqs.(9,13)$ and shown on Figure $12$.}
          \label{tab2}
        \end{center}
      \end{table}

\begin{figure}[h!]
    \begin{center}
   \includegraphics[height=6.40cm,width=10.0cm]{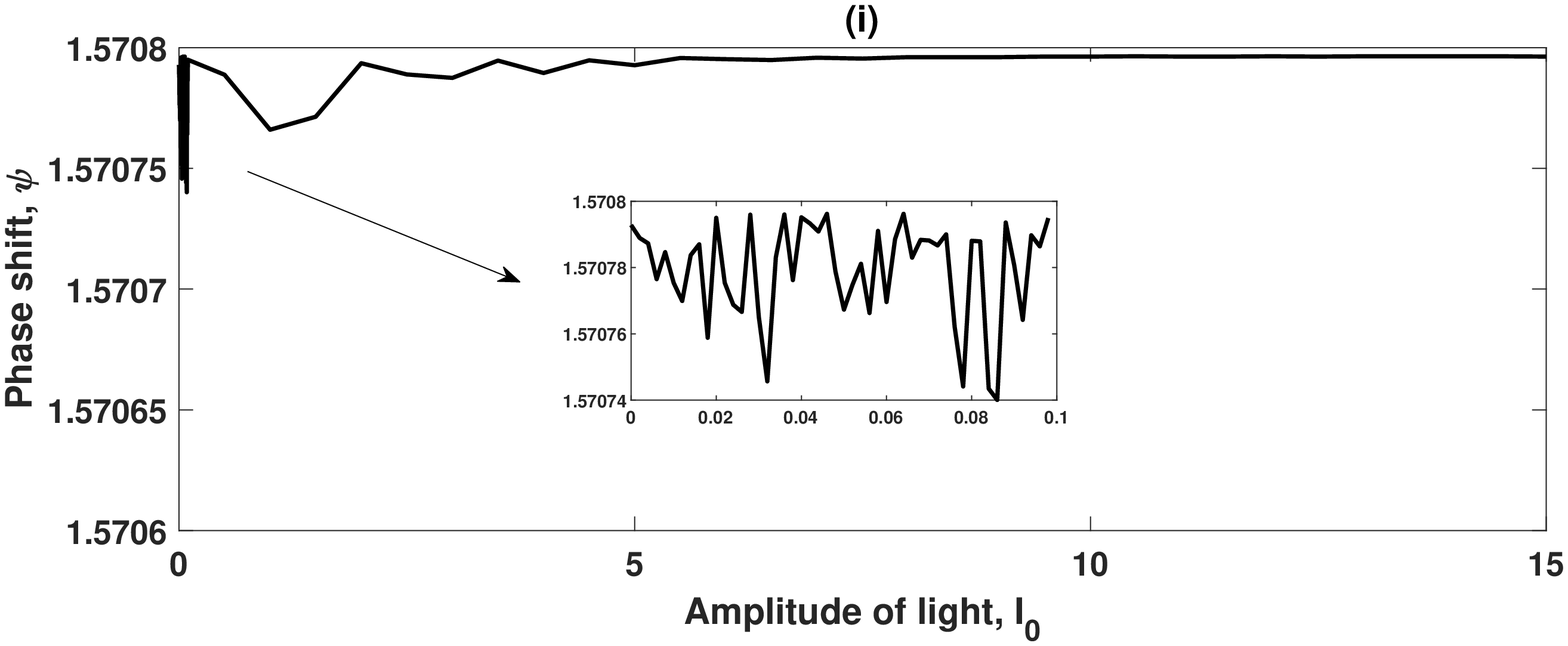}
   \includegraphics[height=6.40cm,width=10.0cm]{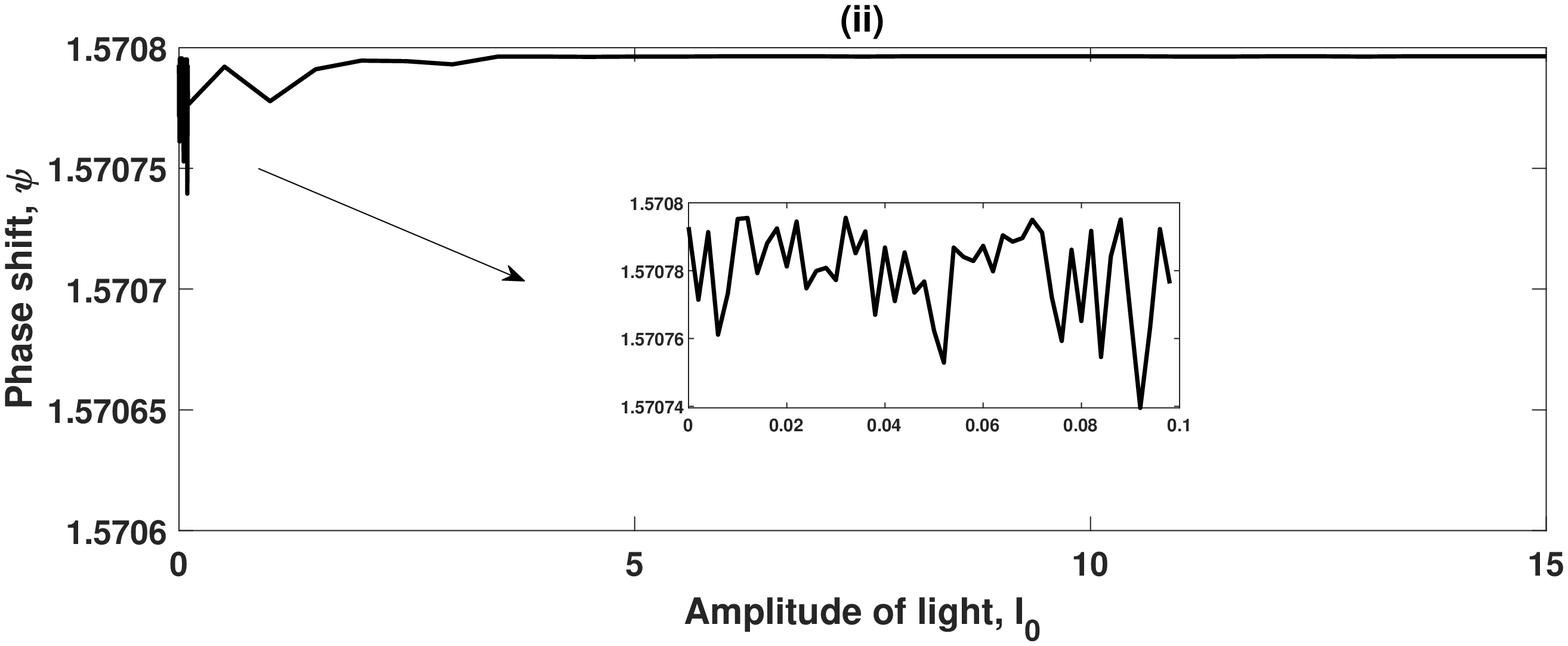}
    \end{center}
   \caption{\it Variation of the phase of circadian oscillations
    versus the amplitude $I_{0}$ of the light for several different
    values of the daylight  $ D_{L} $.   (i) $D_{L}= 10h $ and (ii) $D_{L}= 16h$. The other parameters are defined in Fig. \ref{fig3} and results are obtained using Eqs.(9,13).}
   \label{fig11}
   \end{figure}

By analyzing the phase shift between the light and the mvdP oscillator, let us introduce the following parameter $$ \psi (t) = \arctan\left(\frac{x(t)}{\dot{x}(t)}\right).$$
Figure \ref{fig11} shows its variation as a function of the intensity $I_0$ of the light.
One notices that the phase remains constant, independent of the season, and it is located around the value $\psi = \frac{\pi}{2}.$


\subsubsection{Case of nonzero initial phase of the light: $\phi\neq 0$}

\begin{figure}
\begin{center}
\includegraphics[height=6.40cm,width=10.0cm]{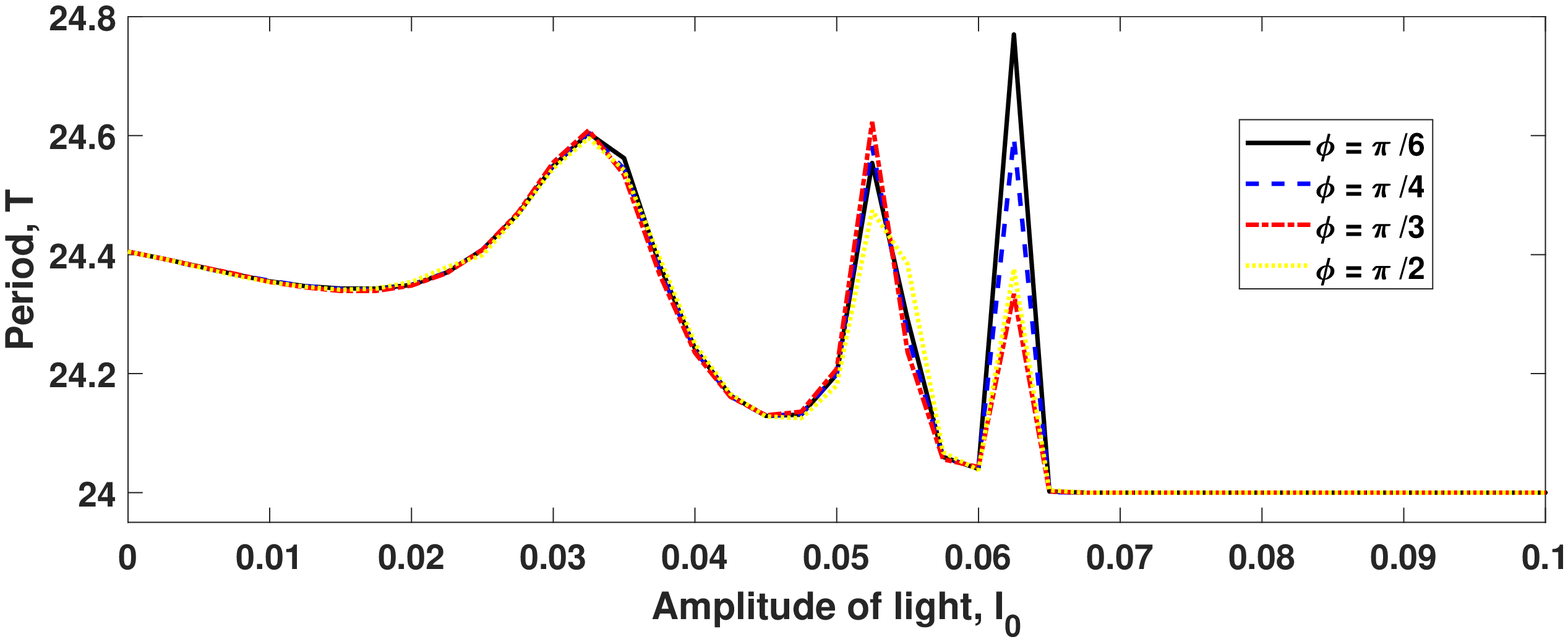}
\caption{\it Effects of the phase $\phi$ on the variation of the period $T$ of circadian oscillations
 versus the amplitude   of light $I_{0}$ for $ D_{L} = 12h$. The other parameters used are defined in figure \ref{fig3} and results are obtained using Eqs.(9,13).
}
\label{fig14}
\end{center}
\end{figure}

\begin{figure}
\begin{center}
\includegraphics[height=6.4cm,width=10.0cm]{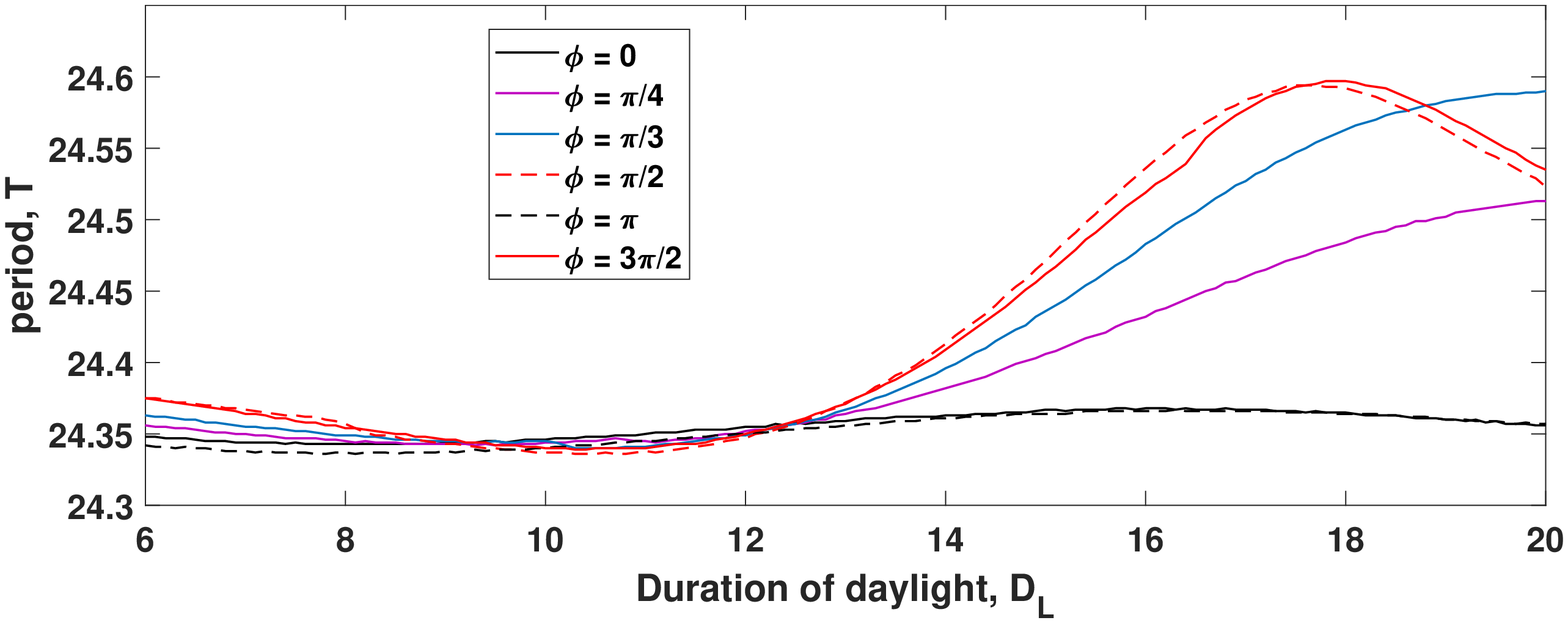}
\caption{\it Effects of the phase $\phi$ on the variation of the period $T$ of circadian
oscillations versus the duration of daylight $D_{L}$ for some different values of $\phi$ with a small value of $ I_{0} = 0.02 $.The other parameters are defined
 in Fig. \ref{fig12} and results are obtained using Eqs.(9,13).
}
\label{fig15}
\end{center}
\end{figure}

\begin{figure}
\begin{center}
\includegraphics[height=6.40cm,width=10.0cm]{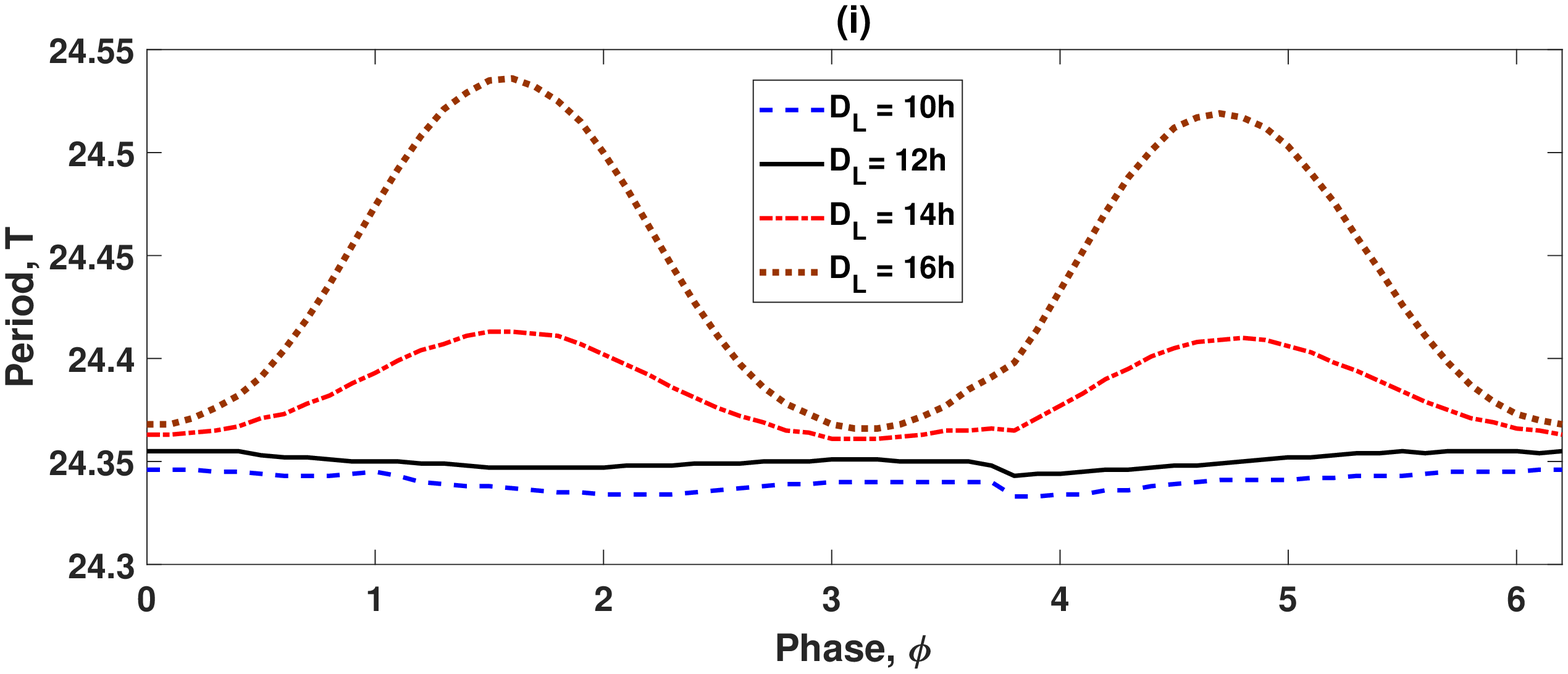}
\includegraphics[height=6.40cm,width=10.0cm]{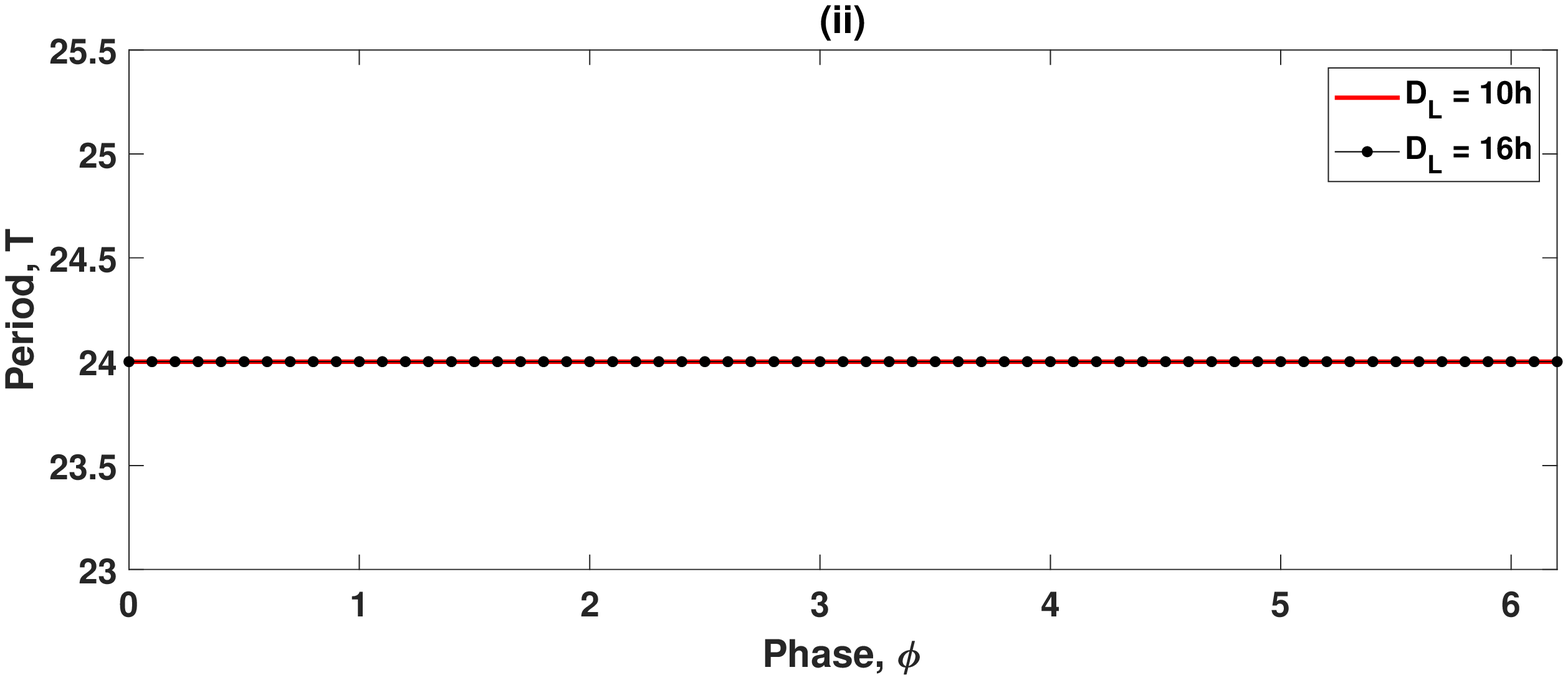}
\caption{\it Effects of the duration of daylight $D_L$ on the
variation of the period $T$ of circadian oscillations versus the phase $\phi$
with  $ I_{0} = 0.02 $ (i) and $ I_{0} = 4 $ (ii). The other parameters are
defined in Fig. \ref{fig12} and results are obtained using Eqs.(9,13).}
\label{fig16}
\end{center}
\end{figure}

\begin{figure}
\begin{center}
\includegraphics[height=6.40cm,width=10.0cm]{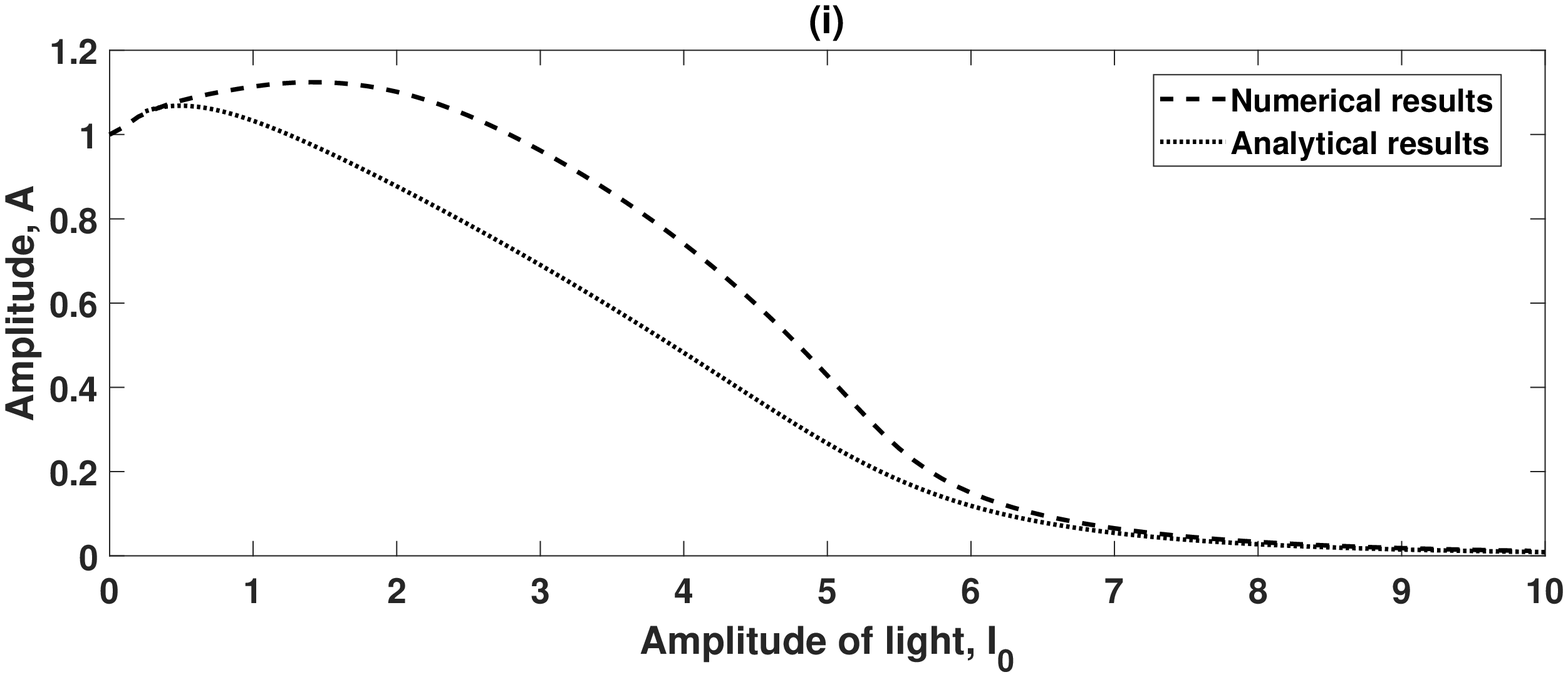}
\includegraphics[height=6.40cm,width=10.0cm]{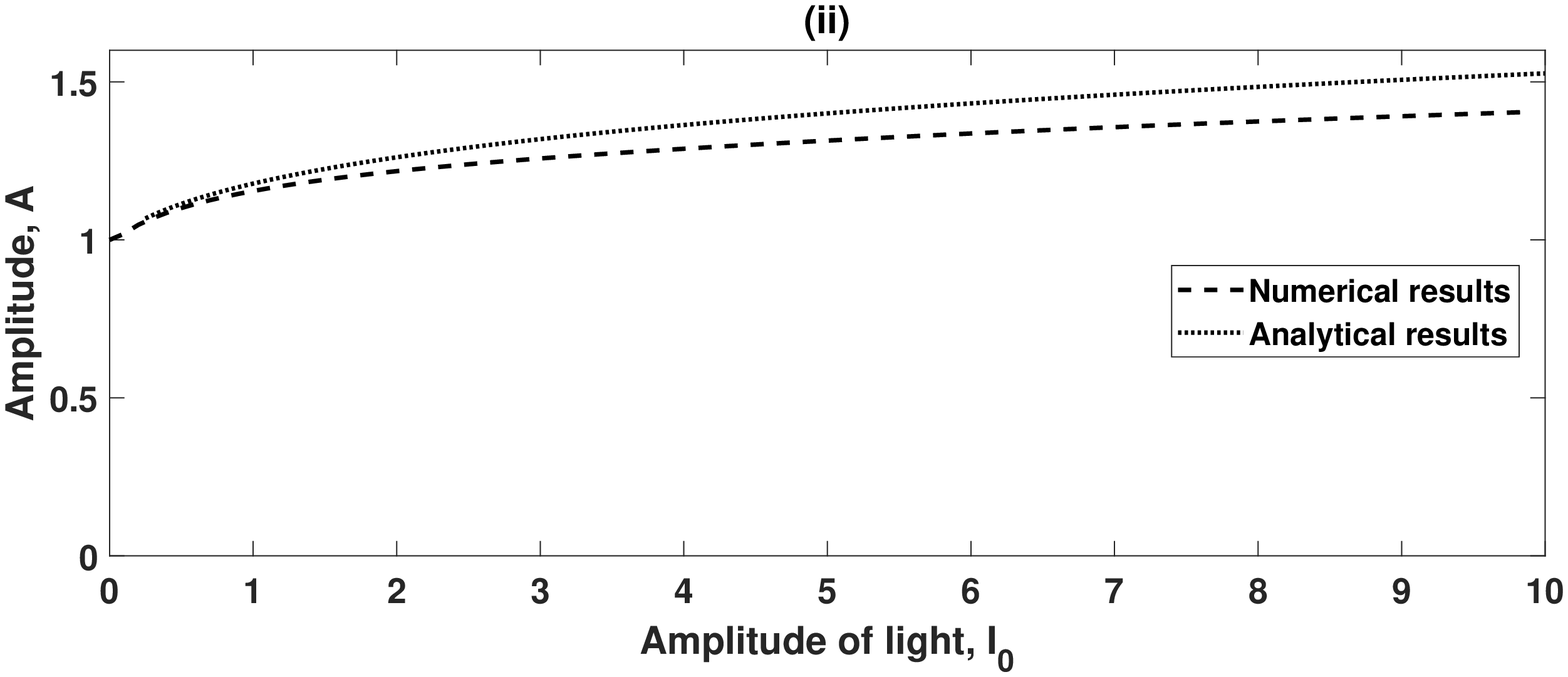}
\caption{\it  Analytical and numerical  amplitude-response versus
the amplitude of light $ I_{0}$ for,  (i)$ \phi = \pi /4 $ and (ii)
$ \phi = \pi /2  $. The duration of daylight used is  $D_{L}= 12h $
and  the other parameters are defined in Fig. \ref{fig12}. 
(Analytical and numerical results are obtained from Eq.(17) and Eqs.(9,13), respectively).}
\label{fig13a}
\end{center}
\end{figure}

\begin{figure}
\begin{center}
\includegraphics[height=6.40cm,width=10.0cm]{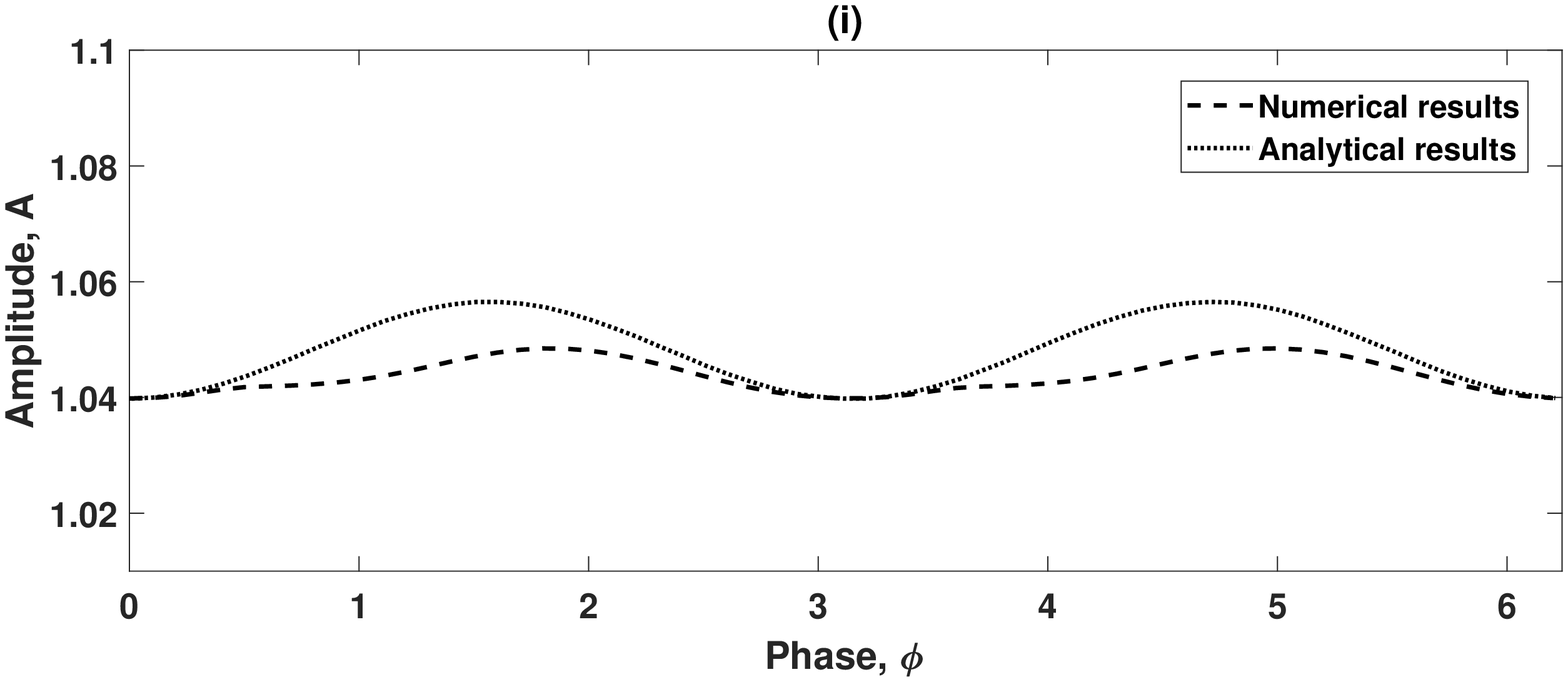}
\includegraphics[height=6.40cm,width=10.0cm]{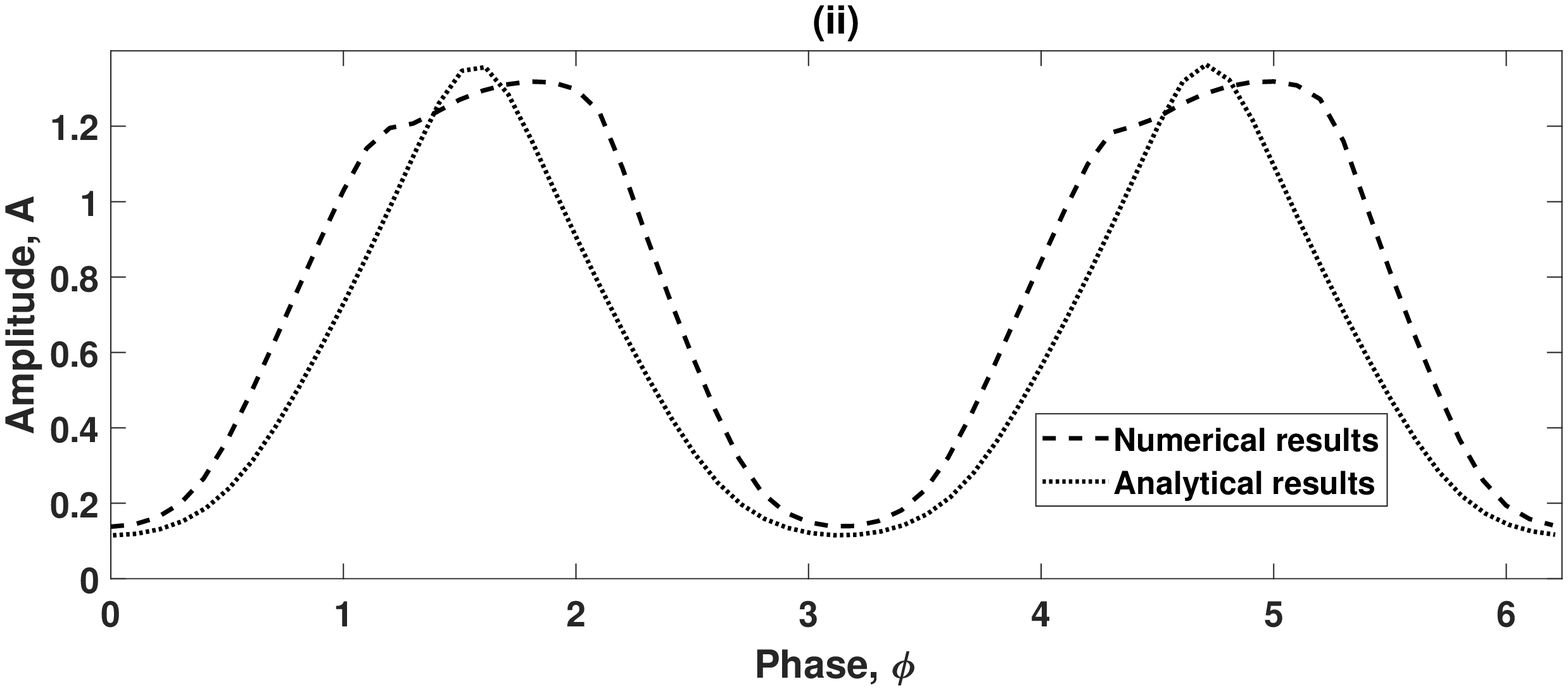}
\caption{\it Analytical and numerical  amplitude-response versus the phase $\phi$ for
 $ I_{0} = 0.2 $ (i) and $ I_{0} = 4 $ (ii). The duration of daylight used is $D_{L}= 12h $
  and the other parameters are defined in Fig. \ref{fig12}.
  (Analytical and numerical results are obtained from Eq.(17) and Eqs.(9,13), respectively).}
\label{fig17a}
\end{center}
\end{figure}

\begin{figure}
\begin{center}
\includegraphics[height=6.40cm,width=10.0cm]{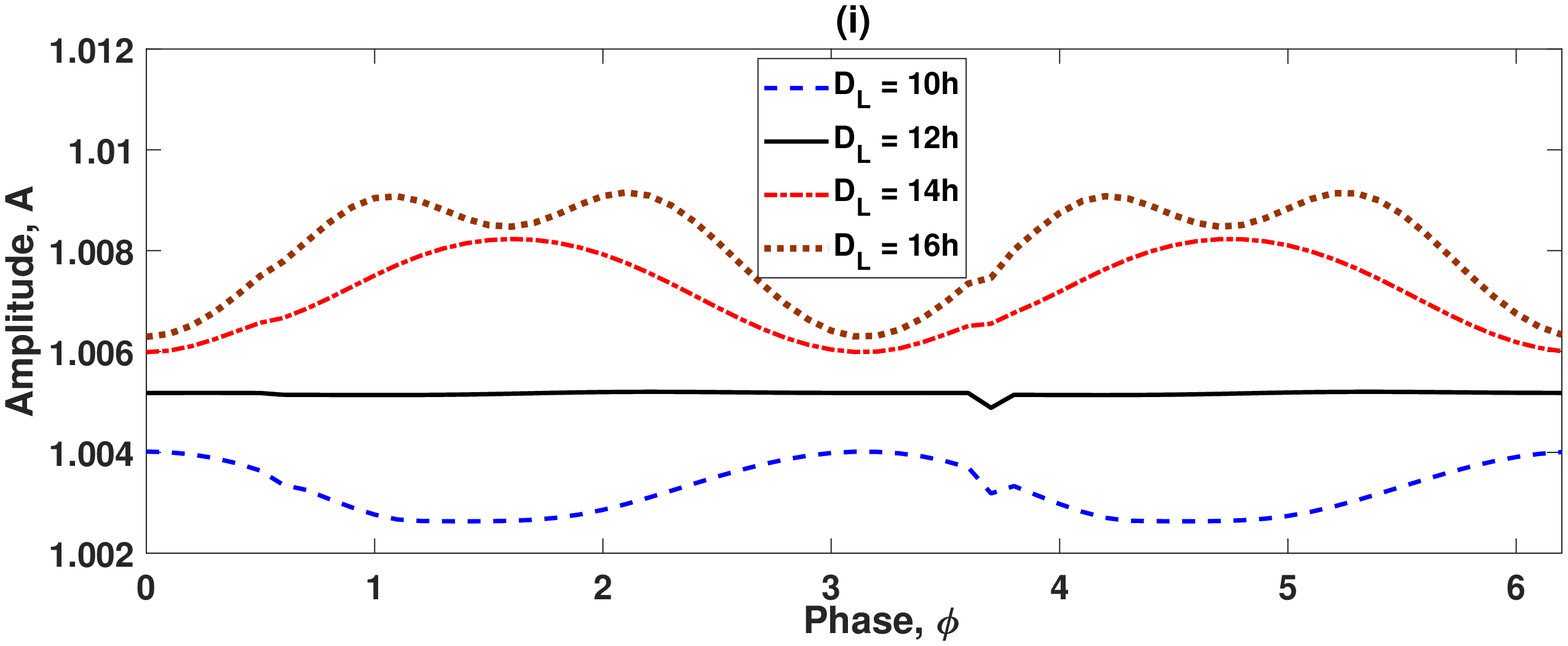}
\includegraphics[height=6.40cm,width=10.0cm]{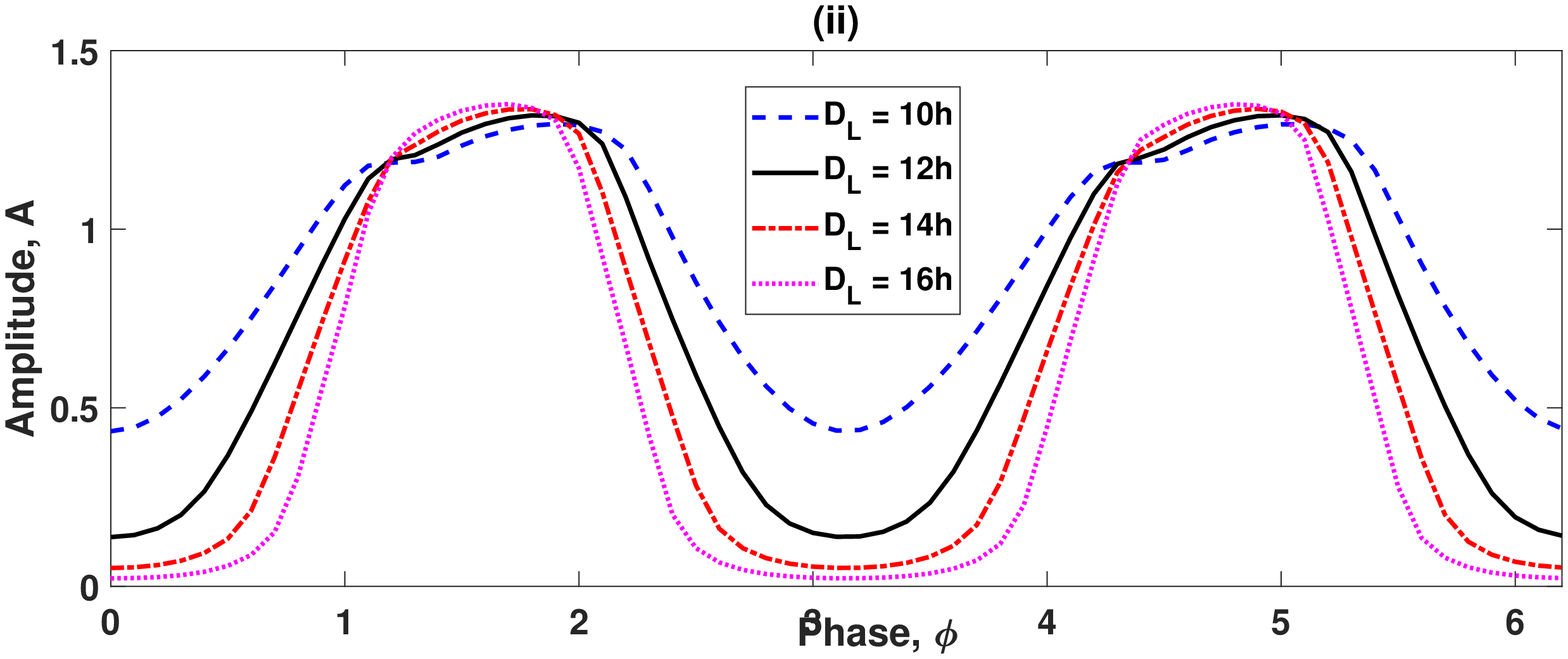}
\caption{\it Effects of the duration of daylight $D_L$ on the
variation of the amplitude-response  versus the phase $\phi$ for
 (i) $I_{0}= 0.02 $,  (ii) $I_{0}= 4$.
 The other parameters are defined
 in Fig. \ref{fig12} and results are obtained using Eqs.(9,13).}
\label{fig17}
\end{center}
\end{figure}

In this subsection, we  consider that there is a non-zero phase difference between the forcing and the oscillator.  In other words, it is a question of studying the wake/sleep cycle in the case where the light acts with a delay or an advance on the circadian stimulator ($ \phi \neq 0 $).
Figs.\ref{fig14} and \ref{fig15} show
the effects of the phase $\phi$ on the variation of the period of circadian oscillations versus
the amplitude of the light $I_0$(see Fig.\ref{fig14}) and
the duration of daylight $D_L$ (see Fig.\ref{fig15}).
It shows that the phase
 shift influences the period of the circadian oscillations if the light
 intensity is low and leaves the period at $24$ hours if the light intensity increases.
 Fig.\ref{fig16}  shows the effects of the duration of daylight $D_L$
 on the variation of the period of circadian oscillations versus phase $\phi$ for two values of
 the intensity $I_0$. It appears in this figure that for very small values of
 the amplitude of light, $I_0$, the period of circadian oscillations oscillates between
  $24.35$ and $24.55$,whereas when $I_0$ becomes large,
  the period of circadian oscillations remains constant
  around $24$ hours. Figures \ref{fig13a} and \ref{fig17a}
   show the analytical and numerical amplitude-response
  versus the amplitude $I_0$ for $\phi=\pi/4$ and $\pi/2$ (see \ref{fig13a})
  and versus $\phi$ for $I_0=0.2$ and $I_0=4.0$
  (see Fig.\ref{fig17a}).
Fig.  \ref{fig17}, we present  the numerical
effects of the daylight duration $D_L$ on the amplitude-response to
 the phase $\phi$ for two different  values of $  I_{0}$.
Fig.19 shows that the phase shift plays a big role on the amplitude of the sleep/wake cycle.  Indeed the amplitude increases when the phase shift approaches towards $ \frac{\pi}{2} $ and decreases when it moves away.

\begin{figure}
\begin{center}
\includegraphics[height=6.40cm,width=10.0cm]{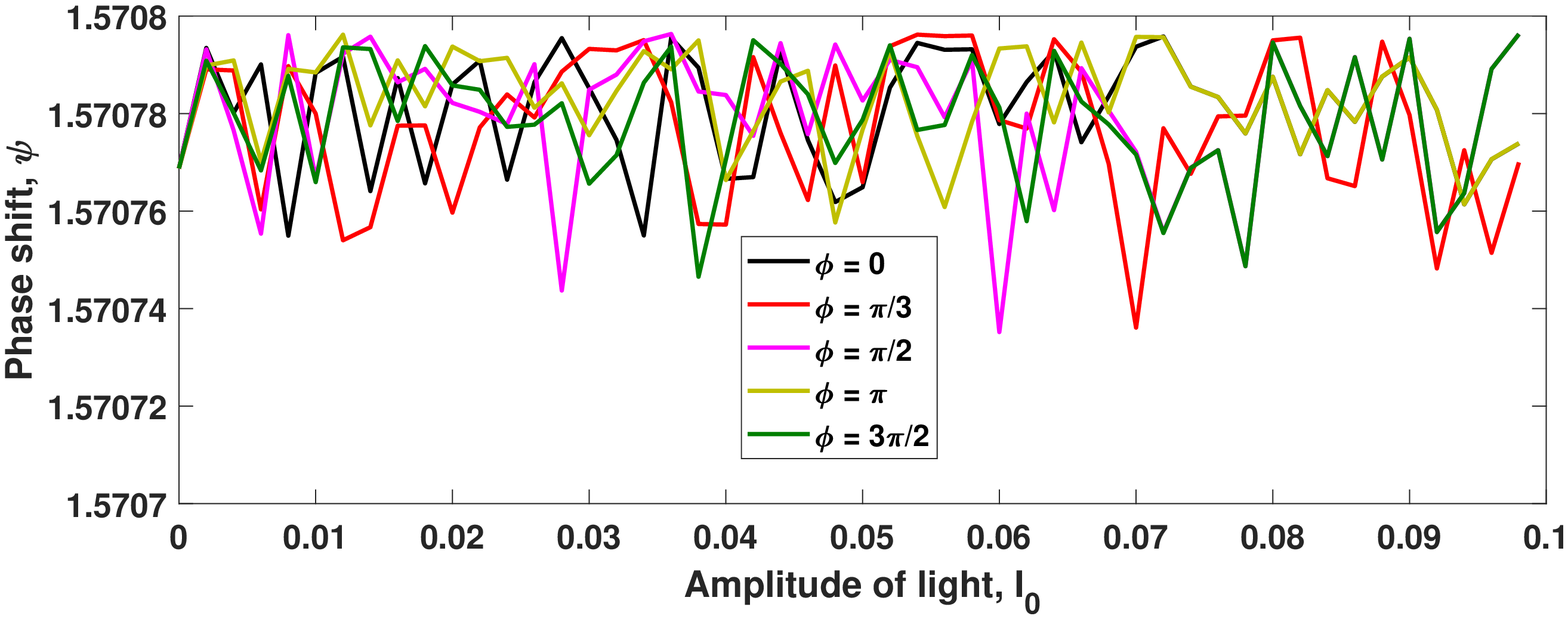}
\caption{\it Variation of the phase of circadian oscillations versus the amplitude $I_{0}$ of the light
    for several different values of  $\phi$ ($D_{L}= 12h $).
      The other parameters are defined in Fig. \ref{fig12} and results are obtained using Eqs.(9,13).}
\label{fig18}
\end{center}
\end{figure}

\begin{figure}
\begin{center}
\includegraphics[height=6.40cm,width=10.0cm]{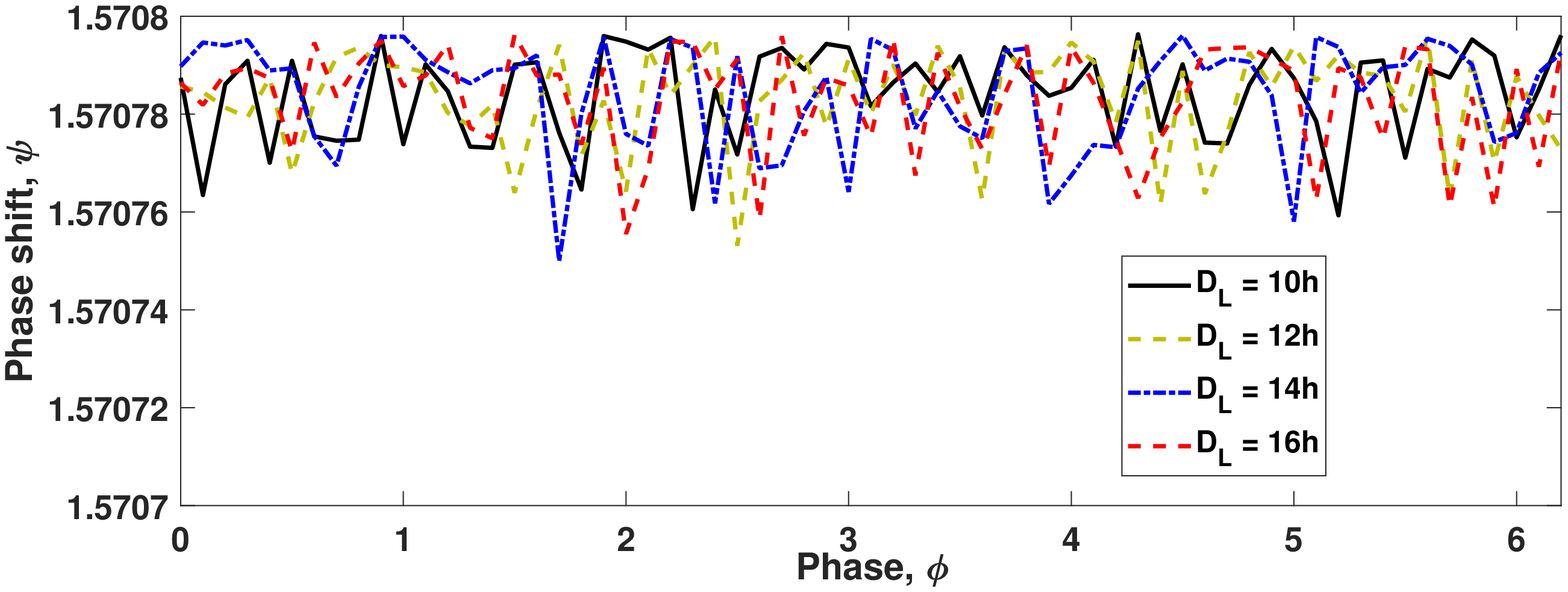}
\caption{\it Variation of the phase of circadian oscillations
versus $\phi$ for  $I_{0}= 0.02 $.
The other parameters are defined in Fig. \ref{fig12}
and results are obtained using Eqs.(9,13).}
\label{fig19}
\end{center}
\end{figure}

\begin{figure}
\begin{center}
\includegraphics[height=6.40cm,width=10.0cm]{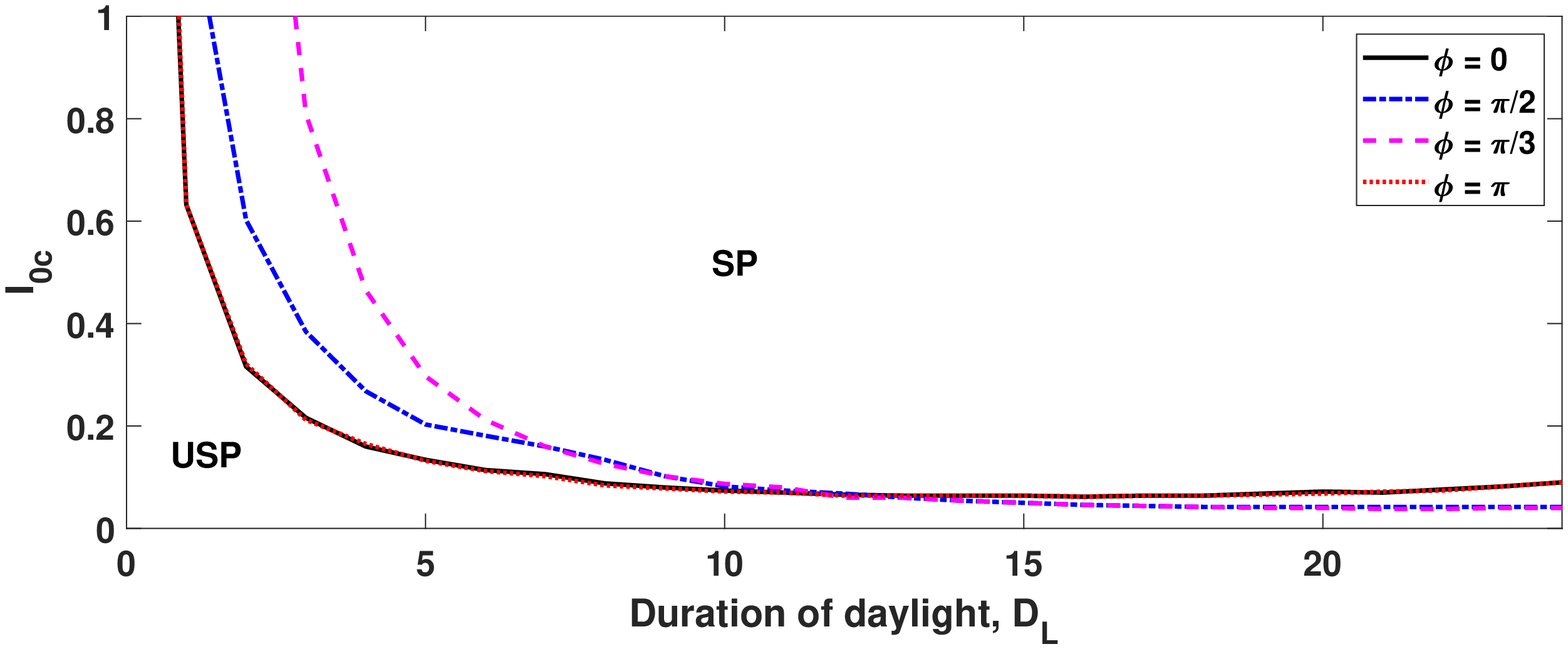}
\caption{\it Effects of the phase $\phi$ on the boundary in the $(I_0,D_L)$ plane between
the region of synchronized period (SP) and the region of unsynchronized period (USP).
The other parameters are defined in Fig. \ref{fig3} and results are obtained using Eqs.(9,13).}
\label{fig20}
\end{center}
\end{figure}

On the other hand according to  Figs.  \ref{fig18} and \ref{fig19}, it appears
 that the phase shift $\phi $ does not really play a big role at
 the level of the phase since it is practically constant
 ($\psi \sim \frac{\pi}{2} $). We can see in
 Fig.  \ref{fig20} that the phase shift does not have a great influence
 on the critical intensity (intensity around which the forcing imposes its frequency on the oscillator).

\begin{figure}
\begin{center}
\includegraphics[height=6.40cm,width=10.0cm]{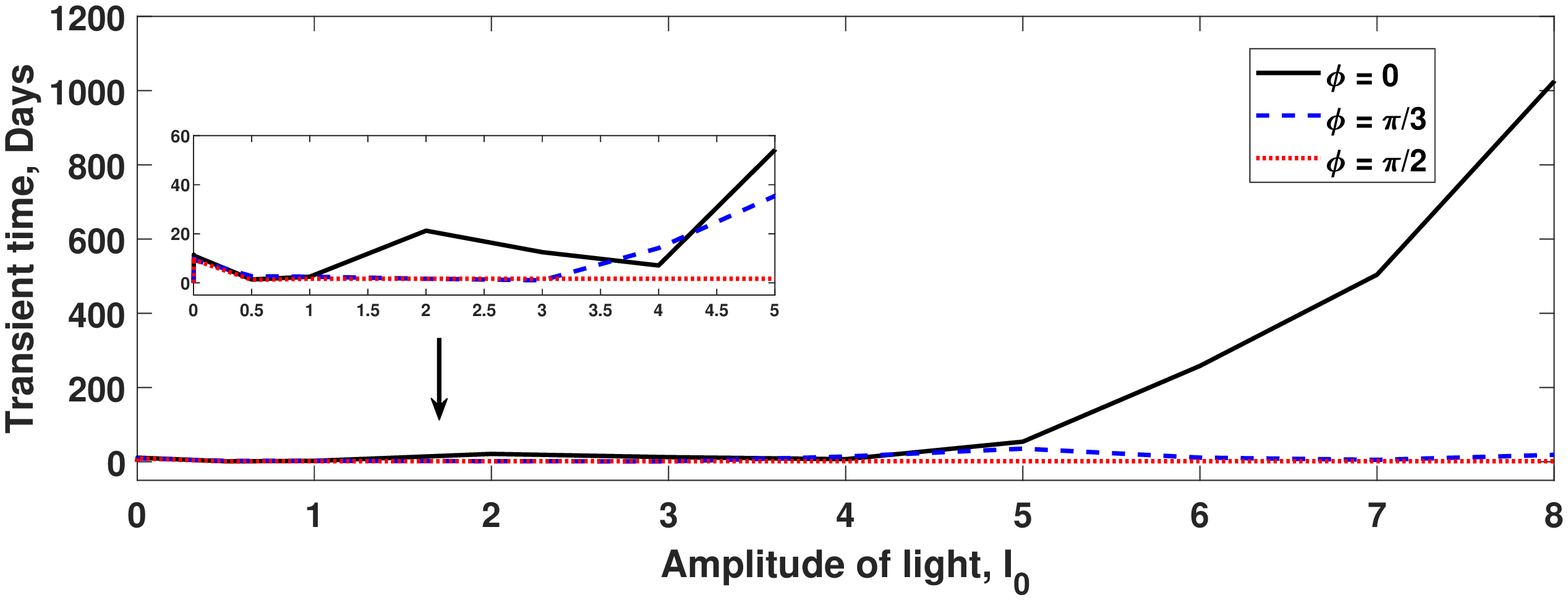}
\caption{\it Effects of the phase $\phi$ on the variation of the transient time versus
$I_{0}$ for $D_{L}= 12h$. The other parameters are defined in Fig. \ref{fig3}
and results are obtained using Eqs.(9,13).}
\label{fig21}
\end{center}
\end{figure}

Figure \ref{fig21} gives more information on the transit time (time at which the  sleep-wake cycle becomes stable after the force is applied). We find that the transit time is relatively small when $  \phi $ approach $\frac{\pi}{2}$, but increases otherwise. Table \ref{tab3} gives us more information on this subject.

\begin{table}[h!]
        \begin{center}
           \begin{tabular}[b]{|c|l|l|l|l|l|l|l|l|l|l|l|l|}
                        \hline
                       \backslashbox{$ D_L$ ; $ \phi$}{$I_{0} $}   & 0.0025 & 0.005  & 0.5 & 1 & 2 & 3 & 4 & 5 & 6 & 7 & 8   \\
                         \hline
12 ; $   \pi/3 $ & 10.63 & 9.38  & 2.70 & 2.50 & 1.67 & 1.04 & 14.16  & 35.41 & 11.25 & 5.21 & 18.75   \\
\hline
12 ; $ \pi/2  $  & 10.41 & 9.37  & 1.25 & 1.67 & 1.67 & 1.67 & 1.67  & 1.67 & 1.67 & 1.67 & 1.67    \\
\hline
16 ; $  \pi/3 $ & 10.41 & 9.38  & 0.83  & 1.25    & 2.29  & 31.25   & 100  & 8.75 & 5.83 & 29.16 & 108.33     \\
\hline
16 ; $  \pi/2 $ & 10.41 & 9.37 & 0.83 & 1.46 & 1.46 & 1.46    & 1.46  & 1.46 & 1.46 & 1.46  &  1.46  \\
                        \hline
                    \end{tabular}
        \caption{\it Transient time(days) for different values of $I_{0} $ and $ DL $ with several values of $ \phi$. These results are obtained by numerical simulations of $Eqs.(9,13)$ and shown on Figure $22$.}
          \label{tab3}
        \end{center}
      \end{table}
The main point is that when the daylight rises earlier or later than usual, it creates a dysfunction in the wake/sleep cycle.  Indeed this can create a variation of the desire to sleep (amplitude of sleep Fig. \ref{fig17}) in particular by increasing or decreasing it according to the delay or the advance.  However,
it does not play a relevant role in the duration of the cycle which varies slightly when the intensity of the external light is very small while it remains constant for high intensity values (Fig. \ref{fig16})  .
 The comparison between analytical and numerical results is acceptable.
Thus, it emerges that the first Fourier component and the ansatz (\ref{eq15}) well describe the main features of the nonlinear oscillator under the effect of day/night light oscillations.
We can therefore conclude that the first component of the Fourier series is the dominant drive.

\subsection{Discussions}

Two more interesting effects occur in the mvdP nonlinear oscillator (\ref{eq10}) that are related to the regularity of the sleep/awake cycle.
First, the time necessary to adapt to light forcing of Fig.23. This is relevant to investigate the effects on the sleep/awake cycle of strong perturbations to the light cycle, as can occur in intercontinental flights in the North or South directions,or in artificial light conditions.
As already shown in Fig. \ref{fig6}, the nonlinear oscillator exhibits a $24$ hour period only if the drive intensity  is large enough  the threshold depends upon the daylight duration.
The effect is illustrated in more details in Fig. \ref{fig7} and
Fig. \ref{fig8}. In the latter figure it is shown the minimal amplitude to obtain synchronization. It is therefore interesting to see that the accommodation time increases with the duration of daylight; indeed this time could take weeks or even months depending on the light intensity.

Another interesting effect might be caused by a sudden change in
 the phase of the oscillator respect to the light cycle, as can 
 occur in intercontinental flights toward East or West \cite{Diekman18}.
Both have a straightforward interpretation in the sleep/awake interpretation 
of the model. The former corresponds to the time necessary to adapt to 
different light conditions, e.g. in intercontinental flights. 
The latter is the minimal amplitude of the light oscillations that cause 
a regular, $24h$ period, sleep/awake cycle.

\section{Conclusions}
\label{conclusions}

A nonlinear oscillator with higher order polynomial dissipation
is at the core of a model for the sleep/awake cycle\cite{Jewett98}.
This oscillator is mathematically very tractable\cite{Yamapi10,Ghosh11,Biswas16,Yamapi12,Yamapi17,Yamapi18},
 and has offered a number of results, from response to sinusoidal drive to effects of noise\cite{Yue12,Guo18}
 (whose effect are not studied here).
In the present work we have shown how the properties of the oscillator
can be interpreted in the case of sleep/awake cycle. Moreover we have
compared the sinusoidal forcing to some more realistic forcing for the light cycle.
The drive can be adapted to reproduce the seasonal changes in the
duration of the daylight, as well as  some particular disorders of the light
stimulus, such as intercontinental flights that cause a sudden change of the phase of the drive.

The simulations show first that a function which is empty for a
certain time (the '' night '') to imitate the behavior of daylight is
 capable, within certain limits, of entraining the oscillator which remains,
 within certain ranges of  parameters, locked to the external drive period
  $24$ hours, the adaptation time depends on the intensity of
   the stimulus and the duration of daylight;   the desire to sleep is
   under the control of each of the stimulus parameters.  Finally, the presence
   of the time difference between the light and the wake-up/sleep cycle slightly
    modifies the period of the system for very small values of light
    intensity
     and locks the system to the period of the
     stimulus ($24$h) at  or beyond -certain $ I_{c} $ values.
     This shift plays a considerable role in the response, that can be interpreted as the desire to sleep.
 The agreement between the analytical results obtained with a treatment of the light features through Fourier analysis  and numerical results is acceptable.

Few words of caution. The proposed oscillator is but a part of more complete mathematical
models that describe the sleep/awake cycle. In addition, it is important
to note that noise, which is unavoidable in biological systems, was not
included in our study.  We wish to underline that a random term should be included in
 future work  more complete models.




\newpage

\newpage

\clearpage

\end{document}